\documentclass[iop, numberedappendix]{emulateapj}		
\usepackage{apjfonts,graphicx}

\newcommand{\zsun}{$Z_\odot$}

\newcommand{\hi}{H\,{\sc i}\rm}
\newcommand{\hii}{H\,{\sc ii}\rm}

\newcommand{\siii}{[S\,{\sc iii}]}

\newcommand{\nii}{[N\,{\sc ii}]}

\newcommand{\oiii}{[O\,{\sc iii}]}
\newcommand{\oiiiit}{[O\,{\it\footnotesize III}]}
\newcommand{\oii}{[O\,{\sc ii}]}

\newcommand{\neiii}{[Ne\,{\sc iii}]}

\newcommand{\op}{O$^{+}$}

\newcommand{\np}{N$^{+}$}

\newcommand{\opp}{O$^{++}$}

\newcommand{\expone}{$^{-1}$}
\newcommand{\gt}{\,$>$\,}
\newcommand{\trgb}{{\sc trgb}}

\newcommand{\zkh}{{\sc zkh}}
\newcommand{\eq}{\,=\,}
\newcommand{\perlum}{{\em P-L}}
\newcommand{\te}{$T_e$}

\newcommand{\hgamma}{H$\gamma$}
\newcommand{\hdelta}{H$\delta$}
\newcommand{\hbeta}{H$\beta$}
\newcommand{\halpha}{H$\alpha$}
\newcommand{\lin}{$\,\lambda$}
\newcommand{\llin}{$\,\lambda\lambda$}

\newcommand{\rtf}{$R_{25}$}

\newcommand{\oh}{12\,+\,log(O/H)}

\newcommand{\rtwothree}{R$_{23}$}
\newcommand{\vs}{vs.}
\newcommand{\ngc}{NGC~4258}

\slugcomment{}

\shorttitle{Abundances in NGC~4258}
\shortauthors{F. Bresolin}

\begin{document}

\title{Revisiting the abundance gradient in the maser host galaxy NGC~4258\footnotemark[1]\\[3mm]} 

\footnotetext[1]{Based on observations obtained at the Gemini Observatory, which is operated by the 
Association of Universities for Research in Astronomy, Inc., under a cooperative agreement 
with the NSF on behalf of the Gemini partnership: the National Science Foundation (United 
States), the Science and Technology Facilities Council (United Kingdom), the 
National Research Council (Canada), CONICYT (Chile), the Australian Research Council (Australia), 
Minist\'{e}rio da Ci\^{e}ncia e Tecnologia (Brazil) 
and Ministerio de Ciencia, Tecnolog\'{i}a e Innovaci\'{o}n Productiva (Argentina). }

\author{Fabio Bresolin} \affil{Institute for Astronomy, 2680 Woodlawn Drive, Honolulu, HI 96822, USA\\ }

\begin{abstract}
New spectroscopic observations of 36  \hii\ regions in \ngc\ obtained with the Gemini telescope are combined with existing data from the literature to measure the radial oxygen abundance gradient in this galaxy. The \oiii\lin4363 auroral line was detected in four of the outermost targets (17 to 22~kpc from the galaxy center), allowing a determination of the electron temperature \te\ of the ionized gas. From the use of different calibrations of the \rtwothree\  abundance indicator an oxygen abundance gradient of approximately $-0.012\pm0.002$~dex\,kpc\expone\ is derived. Such a shallow gradient, combined with the difference in the distance moduli measured from the Cepheid Period-Luminosity relation by Macri et al. between two distinct fields in \ngc, would yield an unrealistically strong effect of metallicity on the Cepheid distances. This strengthens the suggestion that systematic biases might affect the Cepheid distance of the outer field. Evidence for a similar effect in the differential study of M33 by Scowcroft et al.~is presented.
A revision of the transformation between strong-line and \te-based abundances in Cepheid-host galaxies is discussed.
In the \te\ abundance scale, the oxygen abundance of the inner field of \ngc\ is found to be comparable with the LMC value.
\end{abstract}

 
\section{Introduction}

The geometric distance to the spiral galaxy \ngc\ obtained by \citet[$7.2\pm0.5$\,Mpc]{Herrnstein:1999} from the keplerian motion of water masers in its nucleus represents a cornerstone of the extragalactic distance ladder.
The prospect of reducing the uncertainty in the distance from 7\% to 3\% (\citealt{Humphreys:2008})
opens the possibility of adopting \ngc\ as a new anchor of the extragalactic distance scale (\citealt{Riess:2009a}), bypassing the use of the LMC, whose distance remains uncertain at the 5-10\% level.  
Moreover, the availability of a geometric distance for a galaxy located well beyond the Local Group 
offers the opportunity to cross-check the results obtained from primary and secondary distance indicators at distances that are comparable to
those of the calibrators of far-reaching standard candles, such as the type Ia supernova peak brightness.
Distances to \ngc\ that are consistent with the geometrically determined one have been obtained from the Cepheid Period-Luminosity (\perlum) relation (\citealt{Newman:2001}) and the Tip of the Red Giant Branch (\trgb, \citealt{Mager:2008}).
From the Cepheids observed with the Hubble Space Telescope ({\em HST}) by \citet{Macri:2006} various authors, using different treatments of the extinction and metallicity corrections, together with the adoption of different \perlum\ relations, have obtained distance moduli that differ at the 0.1 mag level, but that are still compatible, within the reported uncertainties, with the maser distance  (\citealt{An:2007, 
van-Leeuwen:2007, Benedict:2007, Turner:2010}).

The effect of metallicity on the Cepheid \perlum\ relation is a crucial but still debated issue.
 A persistent discrepancy between theoretical predictions from pulsation models (\citealt{Fiorentino:2007, Bono:2008}) and empirical determinations 
(\citealt{Macri:2006, Scowcroft:2009}) lingers on. In the  {\em HST} Key Project (\citealt{Freedman:2001}) a metallicity correction factor 
$\gamma = \delta\mu_0/\delta\log Z = -0.20$ mag\,dex\expone\ was applied to the galaxy distance moduli, in the sense of making them  larger for metallicities exceeding the LMC value.
The estimated  contribution of the metallicity to the present 
5\% uncertainty on the Hubble constant $H_0$ is about 2.5\% (\citealt{Freedman:2010}).
Current efforts striving for an accuracy better than 5\% on $H_0$ attempt to circumvent this issue by selecting galaxies with similar metal content (\citealt{Riess:2009b}).\smallskip

Given the importance of \ngc\ in current investigations of the extragalactic distance scale, in this paper  the oxygen abundance of its interstellar medium (a proxy for the Cepheid metallicity) is revisited. One of the initial goals of this new study was to determine oxygen abundances from direct electron temperature (\te) determinations in \hii\ regions. This requires deep integrations with large aperture telescopes, in order to detect the weak \oiii\lin4363 auroral line, which provides, in combination with the much stronger \oiii\llin4959,\,5007 nebular lines, the crucial temperature diagnostic for ionized nebulae.
Despite the fact that \ngc\ has been the subject of previous abundance studies based on its \hii\ region population, the \lin4363 line has remained unobserved until the present time. Secondly, it was felt that an investigation based on new spectroscopic data was warranted in order to verify the recent determination of the metallicity effect on the \perlum\ relation obtained by \citet{Macri:2006} from a differential study of Cepheids located in two fields at different distances from the galaxy center (and therefore different metal content). This study adopted metallicities determined from  the calibration of the \rtwothree\ strong-line indicator by \citet{Zaritsky:1994}, applied to \hii\ region  emission-line data obtained by the same authors and \citet{Oey:1993}. However, a recent reanalysis of these data by \citet{Bono:2008} suggested that the abundance gradient might need to be revised. Lastly, during the course of this work some inconsistencies were found in the \te-based abundance scale adopted in several published studies of extragalactic Cepheids (\citealt{Sakai:2004,Saha:2006}), and these will also be addressed for the benefit of future studies.

It is important to keep in mind that the determination of extragalactic chemical abundances from nebular spectroscopy is still afflicted by large systematic uncertainties, despite the maturity of the field (see \citealt{Bresolin:2008} for a recent review). A well-established discrepancy exists between abundances obtained from the classical, {\em direct} \te-based method (\citealt{Menzel:1941}) and the theoretical predictions of photoionization model grids (\citealt{McGaugh:1991}). In addition, the study of metal {\em recombination} lines provides oxygen abundances that lie  between the direct (from {\em collisionally excited} lines) and the theoretical values (\citealt{Garcia-Rojas:2007}). Generally, it is believed that differential analysis methods are unaffected by these systematic differences.
However, in the case of the \te\ \vs\ strong-line methods in the  high metallicity regime (around the solar oxygen abundance \oh$_\odot$\eq8.69, \citealt{Asplund:2009}) this conclusion is still based on a limited number of experimental data points.

\section{Observations and data reduction}
New spectra of \hii\ regions in \ngc\ were obtained with the Gemini Multi-Object Spectrograph (GMOS, \citealt{Hook:2004}) at the Gemini North facility on Mauna Kea. The targets were selected from \halpha\ narrow-band  images of two 5\farcm5$\times$5\farcm5 GMOS fields, centered approximately 2\farcm3 NW and 6\farcm5  SE of the center of the galaxy, obtained between  January and March 2010.
The spectroscopic data were acquired in queue mode between March and June 2010, using two multi-object masks, one per field, with 1\farcs5-wide slitlets.
The seeing conditions were around 0\farcs8 during the 
observations of the NW field, and around 0\farcs5 for the SE field.
In order to minimize the effects of the atmospheric differential refraction the data were acquired at airmasses smaller than 1.20.
Three 3000\,s exposures were secured for each of the two fields using the B600 grating, which provided spectra covering at least
the 3500-5100\,\AA\ wavelength range at a spectral resolution of $\sim$5.5\,\AA. For some of the targets, depending on their spatial distribution, the spectral coverage extended up to $\sim$5900\,\AA. 

{\sc iraf}\footnote{{\sc iraf} is distributed by the National Optical Astronomy Observatories, which are operated by the Association of Universities for Research in Astronomy, Inc., under cooperative agreement with the National Science Foundation.}
routines contained in the {\tt\small gemini/gmos} package were used  for cosmic ray rejection, electronic bias subtraction, flat field correction and wavelength calibration of the raw data frames. Observations of the spectrophotometric standard Feige 34 yielded the flux calibration.
Thanks to the similarity in airmass and atmospheric conditions between the three different frames obtained for each individual target, the 
emission line intensities of the single extracted spectra are  very consistent with each other.
The working version of the spectra was then obtained by averaging the three spectra corresponding to each individual slit.

\begin{figure*}
\medskip
\center \includegraphics[width=0.8\textwidth]{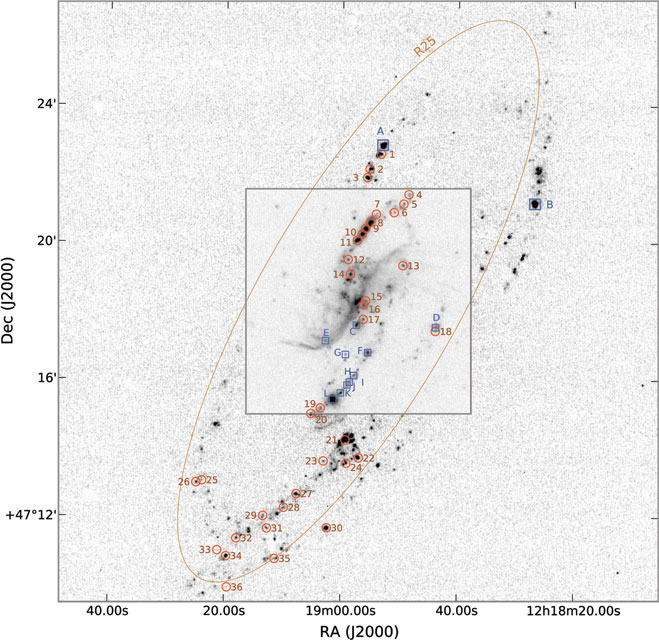}\medskip
\caption{Location of the \hii\/ regions studied in this work on a narrow-band \halpha\/ image of NGC~4258 (courtesy L. van Zee).
The \hii\/ regions are numbered in order of decreasing declination. Letters identify objects drawn from the literature.
The ellipse represents the location of the projected \rtf\/ radius (\rtf\,=\,9\farcm31). The inner and outer parts of the galaxy are shown with different 
stretchings, in order to facilitate the identification of the targets. \label{fig:image} \\}
\end{figure*}

The final \hii\ region sample comprises 36 objects, equally divided between the NW and the SE fields. Their celestial coordinates are summarized in Table~\ref{table:sample} (where objects are listed in order of decreasing declination), together with their galactocentric distances in kpc, deprojected using the geometric parameters in Table~\ref{table:parameters}. Fig.~\ref{fig:image} shows the location of the spectroscopic targets on 
an \halpha\ image of \ngc\ (circles numbered 1 to 36). It can be seen that all but three of the objects lie within the \rtf\ radius (i.e.~within the ellipse in the figure). The \hii\ regions observed in the NW field sample a portion of the northern inner spiral arm of the galaxy, together with a few objects close to the galaxy center and in the northern outer arm. The \hii\ regions in the SE field are mostly located in the southern outer arm (a description of the complex spiral morphology of \ngc\ is given by \citealt{Courtes:1993}).


\begin{deluxetable}{lc}
\tablecolumns{2}
\tablewidth{0pt}
\tablecaption{NGC~4258: adopted parameters\label{table:parameters}}

\tablehead{
\colhead{\phantom{}Parameter\phantom{aaaaaaaaaaaaaaaaaaaaaaaaaaaaaaa}}	     &
\colhead{\phantom{}Value\phantom{}}       }
\startdata
\\[-2mm]
R.A. (J2000.0)\dotfill		& 	12~18~57.50\tablenotemark{a} 				\\
Decl. (J2000.0)\dotfill		&	47~18~14.3\tablenotemark{a}			\\
Morphological type\dotfill					&	SAB(s)bc\tablenotemark{b}					\\
Distance\dotfill				&	7.2 Mpc\tablenotemark{c}					\\
\rtf	\dotfill					&	9\farcm31\tablenotemark{b} (19.50~kpc)				\\		
Inclination\dotfill			&	72$^\circ$	\tablenotemark{d}				\\					
Position angle of major axis\dotfill		&	150$^\circ$	\tablenotemark{d}	\\							
B$_T^0$\dotfill						&	8.53	\tablenotemark{b}			\\				
M$_B^0$\dotfill						&	$-$20.76		
\enddata 
\tablerefs{(a) \citet{Herrnstein:2005}; (b) \citet{de-Vaucouleurs:1991}; (c) \citet{Herrnstein:1999}; (d) \citet{van-Albada:1980}.\\ }
\end{deluxetable}

\begin{deluxetable}{ccccc}
\tabletypesize{\scriptsize}
\tablecolumns{5}
\tablecaption{Observed H\,\scriptsize II \small region sample\label{table:sample}}

\tablehead{
\colhead{\phantom{aaa}ID\phantom{aaa}}	     &
\colhead{\phantom{}R.A.\phantom{}}	 &
\colhead{\phantom{}Decl.\phantom{}}  &
\colhead{\phantom{}R\phantom{}}	 &
\colhead{\phantom{}other ID\phantom{}}	 \\[0.5mm]
\colhead{}       &
\colhead{(J2000.0)}   &
\colhead{(J2000.0)}   &
\colhead{(kpc)}   &
\colhead{} \\[1mm]
\colhead{(1)}	&
\colhead{(2)}	&
\colhead{(3)}	&
\colhead{(4)}	&
\colhead{(5)}	}
\startdata
\\[-2mm]
1\dotfill &  	12~18~53.31	&	47~22~35.5	&  13.66	&	\\
2\dotfill &  	12~18~55.26	&	47~22~09.5	&  13.36	&	\\
3\dotfill &  	12~18~55.69	&	47~21~54.3	&  12.70	&	\\
4\dotfill &  	12~18~48.52	&	47~21~24.8	&  7.58	&	\\
5\dotfill &  	12~18~49.39	&	47~21~08.2	&   6.92	&	\\
6\dotfill &  	12~18~51.05	&	47~20~53.3	&   6.48	&	\\
7\dotfill &  	12~18~54.13	&	47~20~50.3	&   7.62	&	\\
8\dotfill &  	12~18~55.20	&	47~20~35.5	&   7.36	&	34C, ($-032 +141$), B3\\
9\dotfill &  	12~18~55.94	&	47~20~25.5	&   7.23	&	39C, ($-025 +130$), B4\\
10\dotfill &  	12~18~56.55	&	47~20~15.2	&   7.01	&	44C, ($-019 +121$), B5\\
11\dotfill &  	12~18~57.37	&	47~20~04.3	&   6.94	&	52C, ($-010 +110$)\\
12\dotfill &  	12~18~58.93	&	47~19~30.6	&   6.09	&	61C\\
13\dotfill &  	12~18~49.51	&	47~19~20.4	&   5.43	&	18C\\
14\dotfill &  	12~18~58.52	&	47~19~05.0	&   4.10	&	56C\\
15\dotfill &  	12~18~55.87	&	47~18~18.7	&   1.43	&	38C\\
16\dotfill &  	12~18~56.19	&	47~18~11.2	&   1.48	&	42C\\
17\dotfill &  	12~18~56.28	&	47~17~46.0	&   2.89	&	43C\\
18\dotfill &  	12~18~43.86	&	47~17~25.7	&  16.35	&	3C\\
19\dotfill &  	12~19~03.56	&	47~15~10.6	&   7.93	&	82Ca, ($+039 -185$)\\
20\dotfill &  	12~19~05.18	&	47~15~00.8	&   7.93	&	83Ca\\
21\dotfill &  	12~18~59.29	&	47~14~14.8	&  13.97	&	7Sa\\
22\dotfill &  	12~18~56.96	&	47~13~43.3	&  17.79	&	1Sa\\
23\dotfill &  	12~19~03.01	&	47~13~37.6	&  13.79	&	\\
24\dotfill &  	12~18~59.09	&	47~13~33.8	&  16.74	&	\\
25\dotfill &  	12~19~23.88	&	47~13~03.2	&  16.55	&	29S\\
26\dotfill &  	12~19~24.87	&	47~12~59.6	&  17.21	&	32S\\
27\dotfill &  	12~19~07.63	&	47~12~40.2	&  14.79	&	2S\\
28\dotfill &  	12~19~09.78	&	47~12~16.0	&  15.28	&	4S\\
29\dotfill &  	12~19~13.34	&	47~12~01.6	&  15.04	&	\\
30\dotfill &  	12~19~02.38	&	47~11~40.7	&  21.56	&	\zkh~8\\
31\dotfill &  	12~19~12.68	&	47~11~39.6	&  16.30	&	\\
32\dotfill &  	12~19~17.94	&	47~11~22.0	&  16.36	&	\\
33\dotfill &  	12~19~21.18 	&  	47~11~01.0   	&  17.33	&	\\
34\dotfill &  	12~19~19.65	&	47~10~50.6	&  17.61	&	16S\\
35\dotfill &  	12~19~11.33	&	47~10~46.4	&  19.72	&	\\
36\dotfill &  	12~19~19.49 	&  	47~09~56.0	&  19.97	&	
\enddata
\tablecomments{Units of right ascension are hours, minutes and seconds, and units of declination
are degrees, arcminutes and arcseconds. Col.~(1): \hii\/ region identification. Col.~(2): Right
ascension. Col.~(3): Declination. Col.~(4): Deprojected galactocentric distance in kpc.
Col.~(5): Main identification from \citet{Courtes:1993}, with additional objects from 
\citet[=\,\zkh]{Zaritsky:1994}, \citet[coordinates in brackets]{Oey:1993} and \citet[=\,B]{Bresolin:1999}.\\}
\end{deluxetable}


\subsection{Line fluxes and comparison with previous work}
The emission  line intensities were measured with the {\tt\small splot} program in {\sc iraf} by integrating the fluxes under the line profiles.
The metal lines \oii\lin3727,  \oiii\llin4959,\,5007 and the \hbeta,  \hgamma\ Balmer lines were present in the spectra of all the 36 targets in Table~\ref{table:sample}. The weaker \neiii\lin3868 and \hdelta\ lines were measured in 19 and 33 \hii\ regions, respectively.
The line intensities were corrected for interstellar reddening by assuming a case B intrinsic \hgamma/\hbeta\ ratio of 0.47 at \te\eq$10^4$\,K and the \citet{Seaton:1979} reddening law. The resulting reddening-corrected emission lines, normalized to \hbeta\,=\,100, are presented in Table~\ref{table:fluxes}.
The line errors reported in the table account for uncertainties in the flat fielding, the flux calibration, the positioning of the continuum level around each line and the extinction coefficient $c$(\hbeta).

\begin{figure}
\medskip
\epsscale{1.15}
\plotone{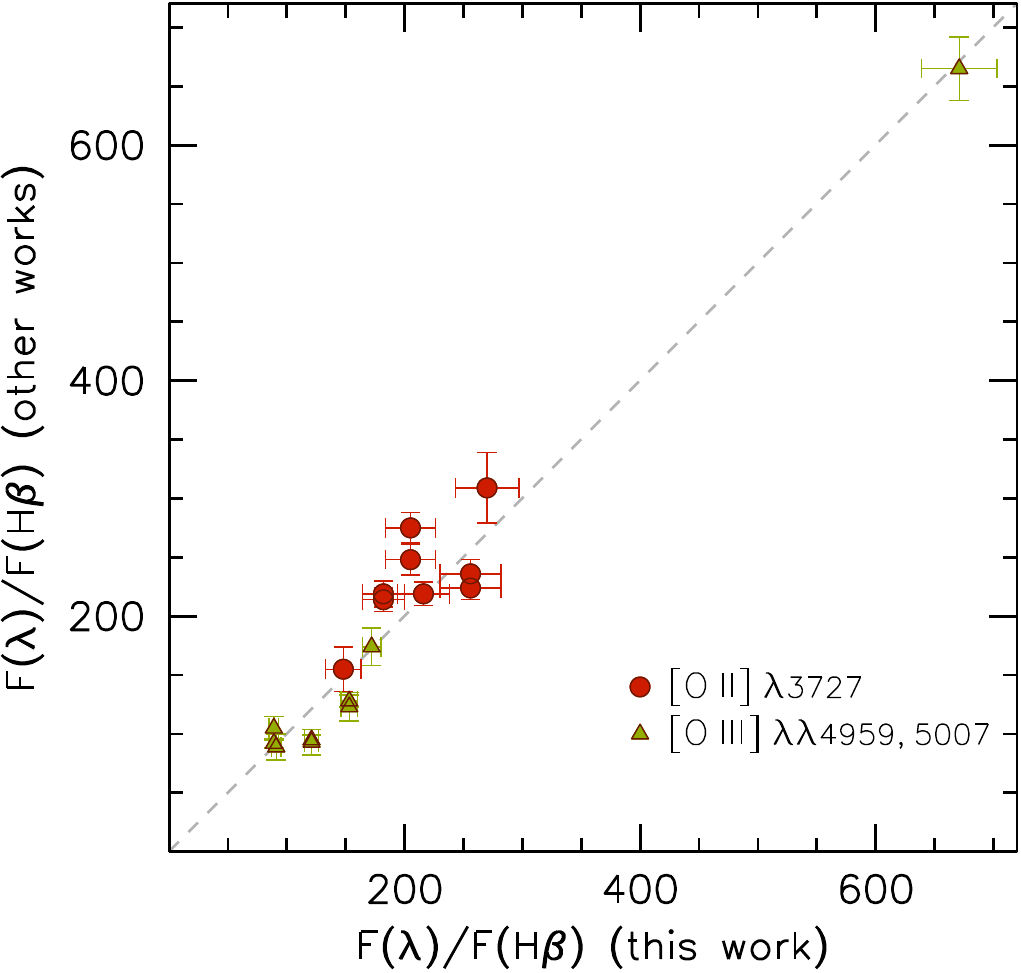}
\caption{Comparison between the \oii\lin3727 and \oiii\llin4959,\,5007 reddening-corrected fluxes  measured in this work and in previous publications. The dashed line represents the one-to-one line.\label{fig:comparison}\\}
\end{figure}

A few of the brightest objects in the sample have already been studied spectroscopically by other authors, in particular six targets are in common with  \citet{Oey:1993}, \citet{Zaritsky:1994} and \citet{Bresolin:1999}, and these are identified in column 5 of Table~\ref{table:sample}.
Fig.~\ref{fig:comparison} compares the \oii\lin3727 (circles) and \oiii\llin4959,\,5007 (triangles) line fluxes measured in the present study with those measured by these other authors. The agreement is found to be excellent: the mean ratio between the two sets is $1.01 \pm 0.04$.

\begin{deluxetable*}{ccccccccccc}
\tabletypesize{\scriptsize}
\tablecolumns{10}
\tablewidth{0pt}
\tablecaption{Reddening-corrected line fluxes and strong-line oxygen abundances\label{table:fluxes}}

\tablehead{
\colhead{\phantom{aaaaa}ID\phantom{aaaaa}}	     &
\colhead{\oii}	 &
\colhead{\neiii}	 &
\colhead{\hi}	 &
\colhead{\oiii}	 &
\colhead{\oiii}	  &
\colhead{log(\rtwothree)}   &
\colhead{F(H$\beta$)} &
\colhead{$c$(\hbeta)} &
 \multicolumn{2}{c}{\oh}      \\[0.5mm]
\colhead{}       &
\colhead{3727}   &
\colhead{3868}   &
\colhead{4101}   &
\colhead{4959}   &
\colhead{5007}   &
\colhead{}   &
\colhead{(erg\,s$^{-1}$\,cm$^{-2}$)}     &
\colhead{(mag)}   &  
\colhead{M91} 	&
\colhead{P05} \\[1mm]
\colhead{(1)}	&
\colhead{(2)}	&
\colhead{(3)}	&
\colhead{(4)}	&
\colhead{(5)}	&
\colhead{(6)}	&
\colhead{(7)}	&
\colhead{(8)}	&
\colhead{(9)}	&
\colhead{(10)}	&
\colhead{(11)}	}
\startdata
\\[-2mm]
 1\dotfill &   273 $\pm$   28 &    13.9 $\pm$  1.8 &      23 $\pm$    2 &     63 $\pm$    4 &    190 $\pm$   11 &   0.72 $\pm$  0.04 &    3.3 $\times 10^{-15}$ &  0.23 &  8.67 $\pm$  0.04 &  8.35 $\pm$  0.05 \\ 
 2\dotfill &   263 $\pm$   27 &    12.5 $\pm$  1.7 &      26 $\pm$    2 &     60 $\pm$    4 &    178 $\pm$   11 &   0.70 $\pm$  0.04 &    3.2 $\times 10^{-15}$ &  0.18 &  8.69 $\pm$  0.04 &  8.37 $\pm$  0.05 \\ 
 3\dotfill &   290 $\pm$   29 &    10.5 $\pm$  1.6 &      26 $\pm$    2 &     58 $\pm$    3 &    166 $\pm$   10 &   0.71 $\pm$  0.04 &    3.3 $\times 10^{-15}$ &  0.12 &  8.67 $\pm$  0.04 &  8.33 $\pm$  0.05 \\ 
 4\dotfill &   248 $\pm$   25 &     \nodata        &      26 $\pm$    2 &     44 $\pm$    3 &    130 $\pm$    8 &   0.62 $\pm$  0.04 &    1.7 $\times 10^{-15}$ &  0.28 &  8.75 $\pm$  0.03 &  8.40 $\pm$  0.05 \\ 
 5\dotfill &   249 $\pm$   25 &     \nodata        &      25 $\pm$    2 &     28 $\pm$    2 &     80 $\pm$    5 &   0.55 $\pm$  0.04 &    4.3 $\times 10^{-15}$ &  0.42 &  8.81 $\pm$  0.03 &  8.36 $\pm$  0.05 \\ 
 6\dotfill &   247 $\pm$   25 &     \nodata        &      20 $\pm$    2 &     21 $\pm$    1 &     65 $\pm$    4 &   0.52 $\pm$  0.04 &    2.6 $\times 10^{-15}$ &  0.17 &  8.82 $\pm$  0.03 &  8.34 $\pm$  0.05 \\ 
 7\dotfill &   219 $\pm$   22 &     \nodata        &      26 $\pm$    2 &     14 $\pm$    1 &     42 $\pm$    3 &   0.44 $\pm$  0.04 &    2.1 $\times 10^{-15}$ &  0.33 &  8.88 $\pm$  0.03 &  8.34 $\pm$  0.04 \\ 
 8\dotfill &   256 $\pm$   26 &     2.9 $\pm$  0.4 &      29 $\pm$    2 &     22 $\pm$    1 &     67 $\pm$    4 &   0.54 $\pm$  0.04 &    2.6 $\times 10^{-14}$ &  0.41 &  8.81 $\pm$  0.03 &  8.33 $\pm$  0.05 \\ 
 9\dotfill &   182 $\pm$   18 &     3.6 $\pm$  0.4 &      24 $\pm$    2 &     38 $\pm$    2 &    115 $\pm$    7 &   0.53 $\pm$  0.04 &    5.5 $\times 10^{-14}$ &  0.27 &  8.84 $\pm$  0.03 &  8.53 $\pm$  0.04 \\ 
10\dotfill &   205 $\pm$   21 &     2.8 $\pm$  0.5 &      24 $\pm$    3 &     30 $\pm$    2 &     91 $\pm$    5 &   0.51 $\pm$  0.04 &    1.7 $\times 10^{-14}$ &  0.50 &  8.84 $\pm$  0.03 &  8.46 $\pm$  0.04 \\ 
11\dotfill &   216 $\pm$   22 &     3.7 $\pm$  0.9 &      25 $\pm$    2 &     23 $\pm$    1 &     68 $\pm$    4 &   0.49 $\pm$  0.04 &    1.3 $\times 10^{-14}$ &  0.20 &  8.86 $\pm$  0.03 &  8.41 $\pm$  0.04 \\ 
12\dotfill &   189 $\pm$   19 &     6.7 $\pm$  1.7 &      24 $\pm$    2 &     69 $\pm$    4 &    205 $\pm$   12 &   0.67 $\pm$  0.03 &    4.7 $\times 10^{-15}$ &  0.44 &  8.74 $\pm$  0.03 &  8.49 $\pm$  0.04 \\ 
13\dotfill &   206 $\pm$   21 &     6.1 $\pm$  0.9 &      27 $\pm$    2 &     40 $\pm$    2 &    119 $\pm$    7 &   0.56 $\pm$  0.04 &    7.8 $\times 10^{-15}$ &  0.05 &  8.81 $\pm$  0.03 &  8.48 $\pm$  0.04 \\ 
14\dotfill &   234 $\pm$   24 &     \nodata        &      24 $\pm$    2 &     29 $\pm$    2 &     87 $\pm$    5 &   0.54 $\pm$  0.04 &    6.1 $\times 10^{-15}$ &  0.35 &  8.81 $\pm$  0.03 &  8.40 $\pm$  0.04 \\ 
15\dotfill &   199 $\pm$   20 &     \nodata        &      25 $\pm$    2 &     24 $\pm$    1 &     72 $\pm$    4 &   0.47 $\pm$  0.04 &    5.1 $\times 10^{-15}$ &  0.00 &  8.87 $\pm$  0.03 &  8.46 $\pm$  0.04 \\ 
16\dotfill &   243 $\pm$   24 &     7.3 $\pm$  1.0 &      26 $\pm$    2 &     50 $\pm$    3 &    148 $\pm$    9 &   0.64 $\pm$  0.04 &    7.9 $\times 10^{-15}$ &  0.37 &  8.74 $\pm$  0.03 &  8.41 $\pm$  0.05 \\ 
17\dotfill &   222 $\pm$   22 &     6.5 $\pm$  0.8 &      26 $\pm$    2 &     44 $\pm$    3 &    128 $\pm$    8 &   0.59 $\pm$  0.04 &    6.5 $\times 10^{-15}$ &  0.16 &  8.78 $\pm$  0.03 &  8.45 $\pm$  0.04 \\ 
18\dotfill &   246 $\pm$   25 &     4.9 $\pm$  0.8 &      28 $\pm$    2 &     38 $\pm$    2 &    112 $\pm$    7 &   0.60 $\pm$  0.04 &    4.2 $\times 10^{-15}$ &  0.00 &  8.78 $\pm$  0.03 &  8.40 $\pm$  0.05 \\ 
19\dotfill &   270 $\pm$   27 &     8.7 $\pm$  1.1 &      26 $\pm$    2 &     43 $\pm$    2 &    129 $\pm$    8 &   0.64 $\pm$  0.04 &    5.5 $\times 10^{-15}$ &  0.39 &  8.73 $\pm$  0.04 &  8.36 $\pm$  0.05 \\ 
20\dotfill &   329 $\pm$   33 &    23.6 $\pm$  2.3 &      30 $\pm$    2 &     74 $\pm$    4 &    223 $\pm$   13 &   0.80 $\pm$  0.04 &    9.0 $\times 10^{-15}$ &  0.90 &  8.58 $\pm$  0.05 &  8.25 $\pm$  0.06 \\ 
21\dotfill &   194 $\pm$   20 &    13.9 $\pm$  2.5 &      26 $\pm$    3 &     56 $\pm$    3 &    176 $\pm$   10 &   0.63 $\pm$  0.03 &    1.9 $\times 10^{-15}$ &  0.66 &  8.76 $\pm$  0.03 &  8.50 $\pm$  0.04 \\ 
22\dotfill &   274 $\pm$   28 &     \nodata        &      26 $\pm$    2 &     26 $\pm$    2 &     74 $\pm$    4 &   0.57 $\pm$  0.04 &    2.1 $\times 10^{-15}$ &  0.20 &  8.78 $\pm$  0.04 &  8.30 $\pm$  0.05 \\ 
23\dotfill &   422 $\pm$   44 &     \nodata        &      24 $\pm$    4 &     48 $\pm$    3 &    142 $\pm$    8 &   0.79 $\pm$  0.04 &    1.0 $\times 10^{-15}$ &  1.14 &  8.56 $\pm$  0.05 &  8.09 $\pm$  0.07 \\ 
24\dotfill &   213 $\pm$   22 &    47.4 $\pm$  4.5 &      25 $\pm$    2 &    157 $\pm$    9 &    465 $\pm$   28 &   0.92 $\pm$  0.03 &    1.6 $\times 10^{-15}$ &  0.26 &  8.50 $\pm$  0.04 &  8.29 $\pm$  0.05 \\ 
25\dotfill &   113 $\pm$   13 &     \nodata        &      23 $\pm$    3 &    104 $\pm$    6 &    303 $\pm$   18 &   0.71 $\pm$  0.03 &    6.0 $\times 10^{-16}$ &  0.35 &  8.73 $\pm$  0.03 &  8.54 $\pm$  0.03 \\ 
26\dotfill &   374 $\pm$   39 &     \nodata        &     \nodata        &      29 $\pm$    2 &     91 $\pm$    5 &   0.70 $\pm$  0.04 &    8.2 $\times 10^{-16}$ &  0.35 &  8.65 $\pm$  0.05 &  8.13 $\pm$  0.06 \\ 
27\dotfill &   355 $\pm$   36 &     \nodata        &      27 $\pm$    3 &     19 $\pm$    1 &     58 $\pm$    3 &   0.64 $\pm$  0.04 &    1.5 $\times 10^{-15}$ &  0.43 &  8.71 $\pm$  0.05 &  8.11 $\pm$  0.06 \\ 
28\dotfill &   286 $\pm$   31 &     \nodata        &      25 $\pm$    4 &     57 $\pm$    4 &    171 $\pm$   10 &   0.71 $\pm$  0.04 &    3.9 $\times 10^{-16}$ &  0.12 &  8.67 $\pm$  0.04 &  8.33 $\pm$  0.06 \\ 
29\dotfill &   337 $\pm$   35 &     \nodata        &      30 $\pm$    4 &     56 $\pm$    3 &    167 $\pm$   10 &   0.75 $\pm$  0.04 &    6.2 $\times 10^{-16}$ &  0.63 &  8.62 $\pm$  0.05 &  8.24 $\pm$  0.06 \\ 
30\dotfill &   149 $\pm$   15 &    47.2 $\pm$  4.5 &      29 $\pm$    2 &    169 $\pm$   10 &    501 $\pm$   30 &   0.91 $\pm$  0.03 &    1.3 $\times 10^{-14}$ &  0.28 &  8.54 $\pm$  0.04 &  8.34 $\pm$  0.04 \\ 
31\dotfill &   314 $\pm$   32 &     \nodata        &      26 $\pm$    3 &     76 $\pm$    4 &    231 $\pm$   14 &   0.79 $\pm$  0.04 &    7.8 $\times 10^{-16}$ &  0.08 &  8.59 $\pm$  0.04 &  8.28 $\pm$  0.06 \\ 
32\dotfill &   287 $\pm$   30 &     \nodata        &     \nodata        &      14 $\pm$    1 &     42 $\pm$    3 &   0.54 $\pm$  0.04 &    8.5 $\times 10^{-16}$ &  0.07 &  8.80 $\pm$  0.04 &  8.20 $\pm$  0.05 \\ 
33\dotfill &   241 $\pm$   27 &     \nodata        &     \nodata        &     118 $\pm$    7 &    353 $\pm$   21 &   0.85 $\pm$  0.03 &    3.2 $\times 10^{-16}$ &  0.08 &  8.56 $\pm$  0.04 &  8.33 $\pm$  0.05 \\ 
34\dotfill &   314 $\pm$   32 &    18.8 $\pm$  1.9 &      26 $\pm$    2 &     80 $\pm$    5 &    237 $\pm$   14 &   0.80 $\pm$  0.03 &    5.5 $\times 10^{-15}$ &  0.20 &  8.58 $\pm$  0.04 &  8.28 $\pm$  0.06 \\ 
35\dotfill &   326 $\pm$   33 &     \nodata        &      23 $\pm$    3 &     52 $\pm$    3 &    157 $\pm$    9 &   0.73 $\pm$  0.04 &    8.0 $\times 10^{-16}$ &  0.00 &  8.64 $\pm$  0.04 &  8.26 $\pm$  0.06 \\ 
36\dotfill &   193 $\pm$   20 &    32.9 $\pm$  4.3 &      25 $\pm$    2 &    169 $\pm$   10 &    492 $\pm$   29 &   0.93 $\pm$  0.03 &    6.0 $\times 10^{-16}$ &  0.00 &  8.50 $\pm$  0.04 &  8.29 $\pm$  0.05 
\enddata
\tablecomments{The line fluxes  are in units of H$\beta$\,=\,100. F(H$\beta$) in column (8) is the measured \hbeta\ flux, uncorrected for extinction.
The oxygen abundances in columns (10) and (11) are obtained with the \rtwothree\ calibrations by 
\citet[=\,M91]{McGaugh:1991} and \citet[=\,P05]{Pilyugin:2005a}, respectively.\\}
\end{deluxetable*}

\section{Oxygen abundances}

\subsection{Electron temperatures and direct abundances}
One of the goals of this new spectroscopic investigation of \ngc\ was to measure, for as many \hii\ regions as possible, the electron temperature of the nebular gas via the standard method that involves the detection of the \oiii\lin4363 auroral line. No previous detection of this line has been reported in the literature for \ngc. In this new study, the \lin4363 line was measured for four \hii\ regions (numbers 24, 30, 34 and 36 in Table~\ref{table:sample}), all of them located at large galactocentric distances and along the southern outer spiral arm. The {\tt\small nebular} package in {\sc iraf} was used to derive electron temperatures (adopting an electron density of 10$^2$\,cm$^{-3}$ and the atomic parameters used by \citealt{Bresolin:2009a}), and the \op, \opp\ ionic abundances. The resulting temperatures and total O/H abundance ratios are presented in Table~\ref{table:auroral}. The electron temperatures and the \oh\ abundances are remarkably similar among the four targets, with mean values of 11400~K and 8.20, respectively. 

\begin{deluxetable}{cccc}
\tabletypesize{\scriptsize}
\tablecolumns{4}
\tablecaption{Direct method abundances\label{table:auroral}}
\tablehead{
\colhead{\phantom{aaaaa}ID\phantom{aaaaa}}	     &
\colhead{\oiii\lin4363}	 &
\colhead{\te (K)}  &
\colhead{12\,+\,log(O/H)}	  \\[1mm]
\colhead{(1)}	&
\colhead{(2)}	&
\colhead{(3)}	&
\colhead{(4)}	}
\startdata
\\[-2mm]
24\dotfill	&	4.5 $\pm$ 0.7   & 11350 $\pm$ 850 		& 8.23 $\pm$ 0.10 \\  
30\dotfill	&	5.0 $\pm$ 0.4	&	11490 $\pm$ 530 	& 8.19 $\pm$ 0.06 \\  
34\dotfill	&	1.9 $\pm$ 0.2	&	10640 $\pm$ 510 	& 8.23 $\pm$ 0.07  \\  
36\dotfill	&	5.6 $\pm$ 1.0	&	12030 $\pm$ 1050 	& 8.16 $\pm$ 0.11 
\enddata
\tablecomments{The reddening-corrected intensity of \oiii\lin4363 in column 2 is in units of H$\beta$\,=\,100.\\}
\end{deluxetable}

\subsection{Strong-line abundances}
For the majority of the \hii\ regions in our sample we need to resort to the use of strong-line abundance indicators, since the auroral lines remain undetected. The difficulties and potential pitfalls of these methods have been described at length in the literature (see \citealt{Bresolin:2009a} for a recent overview) and will not be repeated here. In brief, with these methods the value of the oxygen abundance is related to the strength of metal lines in the optical nebular spectra either via an empirical calibration obtained from \hii\ regions with auroral line detections ('empirical' method) or via the results of theoretical photoionization models. The abundances derived from the different procedures and calibrations vary in a systematic way by large factors, with empirical abundances typically occupying the low end of the range. While this situation is currently unexplained, relative abundances are generally considered to be rather robust. Investigations of additional present-day abundance indicators in galaxies (e.g.~blue supergiants, \citealt{Kudritzki:2008}) are presently being carried out, with the goal of establishing the reliability  of the absolute nebular abundances obtained by the different calibrations. Recent work in the nearby galaxies NGC~300 and M33 has shown that a good agreement is obtained between the supergiant and \hii\ region abundances when the latter are based on the direct, \oiii\lin4363-based method (\citealt{Bresolin:2009a,Bresolin:2010}). For this reason, it is important to be able to anchor the strong-line abundances obtained in more distant galaxies to the values provided by the detection of the \oiii\lin4363 auroral line, as done in the present investigation of \ngc. While relative abundance determinations seem to be, to first order, unaffected by the choice of technique used to measure them (but see last paragraph of Sec.~\ref{sec:tescale}), 
it is also true that it is important to establish the absolute nebular abundances  in order to constrain the chemical evolutionary status of star-forming galaxies, a fact that is often overlooked.

For this study  of \ngc\ it was decided to investigate the abundances obtained from the strong-line indicator \rtwothree\,=\,(\oii\lin\,3727 + \oiii\llin4959,\,5007)/\hbeta\ (\citealt{Pagel:1979}).
Two different calibrations are adopted: the one by \citet[in the analytical form given by \citealt{Kuzio-de-Naray:2004}]{McGaugh:1991}, which is based on a grid of photoionization models, and the one by \citet{Pilyugin:2005a}. The latter was selected as an alternative because it is essentially calibrated from auroral line-based abundances, and thus it should provide results that are in agreement with those obtained from the  \oiii\lin4363 detections. Both calibrations account for the excitation of the ionized gas, via the parameters $y$\eq log(O3/O2) and $P$\eq O3/(O3\,+\,O2), respectively, where O3\eq \oiii\llin4959,\,5007 and 
O2\eq \oii\lin\,3727.

\begin{figure}
\medskip
\epsscale{1.15}
\plotone{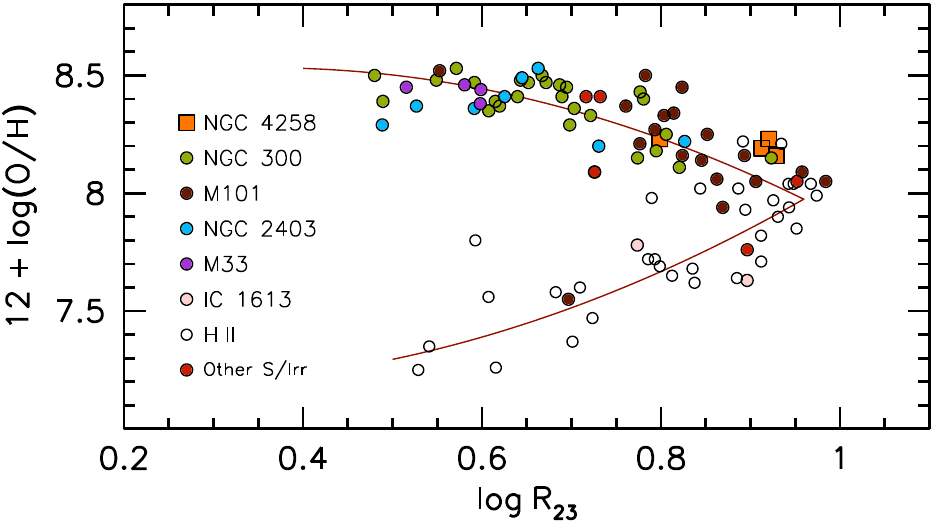}
\caption{Log(\rtwothree) \vs\ \oh\ for a sample of extragalactic \hii\ regions drawn from the literature. The oxygen abundances were all derived using the direct method, i.e.~using the \oiii\lin4363 auroral lines. The different symbols correspond to the galaxies identified in the legend. References for the data are given in the text. The four \hii\ regions in \ngc\ with a measured \oiii\lin4363 line are shown by the orange squares. The curves represent the upper and lower branch of the \rtwothree\ indicator, obtained from a polynomial fit to the data.\\
\label{fig:r23}}
\end{figure}

The double-valued nature of \rtwothree\  represents a major source of uncertainty when it is difficult to ascertain whether the targets fall onto the upper (high O/H) or lower (low O/H) branch of the indicator.
This issue is commonly resolved by considering emission line ratios that are monotonic with O/H, such as \nii/\halpha\ and \nii/\oii.
Our spectra do not cover the region around \halpha, and therefore we can neither use indicators based on the strength of the \nii\lin6583 line (e.g.~N2\,=\,log(\nii/\halpha, \citealt{Pettini:2004}), nor use this information to select the appropriate \rtwothree\ branch. Nevertheless,
it is possible to select the upper branch calibration for the full sample with high confidence, based on the following considerations:

\smallskip

\noindent
{\em (a)} Fig.~\ref{fig:r23} shows the relation between \rtwothree\ and O/H for \hii\ regions in a number of spiral and irregular galaxies, for which the oxygen abundance  was derived from high-quality measurements of the \oiii\lin4363 line and extracted from the following references:\\

\noindent
\begin{tabular}{ll}
NGC~300:						&		\citet{Bresolin:2009a}\\[0.5mm]
M101: 							&		\citet{Kennicutt:2003}\\[0.5mm]
NGC~2403:						&		\citet{Garnett:1997}\\
								&		\citet{Esteban:2009}\\[0.5mm]
M33: 							&	 	\citet{Bresolin:2010}\\
								&		\citet{Esteban:2009}\\[0.5mm]	
IC~1613: 							&		\citet{Bresolin:2006a}\\[0.5mm]
\hii\ galaxies: 						&		\citet{Izotov:1994}\\
								&		\citet{Guseva:2009}\\[0.5mm]
M31, NGC~2363, NGC~1741			&		\citet{Esteban:2009}\\					
NGC~4395, NGC~4861				&	\\
					
\end{tabular}

\medskip
In selecting the spiral galaxy sample preference was given to investigations carried out by the author and his collaborators, in order to ensure consistency in the reduction and data analysis procedures.
The data plotted in Fig.~\ref{fig:r23} clearly delineate the two \rtwothree\ branches, with the turnover point located around \oh\,=\,8.0. Most of the \hii\ galaxies shown here fall below the turnover point and lie on the lower branch. Virtually all of the \hii\ regions in spiral galaxies are located on the upper branch (the notable exception is represented by region SDH~323 in M101, which lies at a very large distance from the galaxy center).
The lines drawn in Fig.~\ref{fig:r23} are polynomial fits to the data, for \oh\ above and below the turnover point, and are shown to guide the eye in identifying the two \rtwothree\ branches. 
The four squares in Fig.~\ref{fig:r23} represent the \hii\ regions in \ngc\ for which a direct abundance was obtained (see Table~\ref{table:auroral}). It can be seen that these objects are placed on the upper branch.
Considering the fact that these \hii\ regions are among the most distant from the center of \ngc, and that galactic abundance gradients are always measured to be monotonic, with abundances increasing toward the central regions, this result strongly suggests that  the remaining \hii\ regions in this galaxy (which have smaller \rtwothree\ values, as shown in column 7 of Table~\ref{table:fluxes})
have larger abundances, and should therefore also be placed on the upper branch.

\medskip
\noindent
{\em (b)} all of the \hii\ regions without \oiii\lin4363 detections have log(\rtwothree)\,$<$\,0.80, i.e.~significantly lower than the turnaround point where the two branches intersect. From Fig.~\ref{fig:r23} we can infer that misplacing a lower branch object onto the upper branch would increase the estimated O/H ratio by approximately 0.5 dex or more. This would easily show up in a plot of the galactocentric abundance gradient, under the assumption that the gradient is monotonic {\em and} without abrupt discontinuities. If we assume upper branch objects only for our sample, the gradient in \ngc\ behaves `normally'. If instead we assume that some of the objects at high \rtwothree\ values belong to the lower branch, we obtain very large discontinuities in the abundance gradient, which can be reasonably excluded for \hii\ regions located within the optical boundaries of normal galaxies.

\medskip
\noindent
{\em (c)} For four of the targets (4, 5, 13 and 34 in Table~\ref{table:fluxes}) the auroral \nii\lin5755 line was within the covered wavelength range and sufficiently strong to allow a good measurement of its intensity. This line, in combination with the much stronger \nii\lin6583, can be used to measure the electron temperature in the low-ionization zone of an \hii\ region, in order to determine, for example, the \np\ and \op\ ionic abundances (\citealt{Bresolin:2004}).
Here we can use it to infer the strength of the \nii\lin6583 line, making a reasonable assumption for the value of the electron temperature. Then, the N2\eq log(\nii\lin6583/\halpha) index can be used in the traditional way to assign the objects to the lower or upper \rtwothree\ branch.
For example, assuming \te\,=\,10,000~K (\te\,=\,10,600~K for object 34, as derived from the \oiii\lin4363 line), from the {\tt\small temden} routine in {\sc iraf} we obtain N2\eq $-0.72\pm0.08, -0.61\pm0.05, -0.77\pm0.05, {\rm and} -0.94\pm0.06$ for the four objects, respectively. It is generally adopted that for N2\,$>$\,$-1.2$, the upper branch solution of \rtwothree\ should be preferred (e.g.~\citealt{Kewley:2008}). The boundary between upper and lower branch could be safely lowered to N2\,$\simeq$\,$-1.4$, if we consider that the upper and lower branches intersect around \oh\,=\,8.0, when the abundances are determined empirically (see Fig.~\ref{fig:r23}). A plot of empirical (\oiii\lin4363-based) abundances \vs\ N2 shows that this value corresponds to N2\,=\,$-1.5$ (\citealt{Pettini:2004}). In any case, the N2 values we estimated above place the targets safely  in the upper branch. Only once we approach a \te\ value of 14,000~K would the N2 estimates be below $-1.2$. 
However, such high temperatures can be reasonably excluded by pointing out that the directly measured electron temperature values, approximately between 11,000~K and 12,000~K, are found for objects located at the largest galactocentric distances in \ngc, while the electron temperature is expected to decrease towards smaller radii (\citealt{Bresolin:2009a}).\\

In conclusion, there are strong arguments to favor the selection of upper branch calibrations of \rtwothree\ for the whole \hii\ region sample analyzed in this paper. 

\section{The oxygen abundance gradient in \ngc}\label{section:oxygen}
The adoption of the \rtwothree\ upper branch, justified in the previous section, yields the oxygen abundances presented in columns (10) and (11) of Table~\ref{table:fluxes}, obtained with the \citet[=\,M91]{McGaugh:1991} and \citet[=\,P05]{Pilyugin:2005a} calibrations, respectively.
For the derivation of the abundance gradient the line fluxes of 12 additional \hii\ regions published by \citet{Oey:1993}, \citet{Zaritsky:1994} and \citet{Diaz:2000} were added to our sample. These objects are identified with letters in Fig.~\ref{fig:image}, and their positions and galactocentric distances can be found in Table~\ref{table:literature}. When the same target was observed by these authors, the average \oii\ and \oiii\ line fluxes were computed. For objects from our sample that are in common with these authors (see Table~\ref{table:sample} and Fig.~\ref{fig:comparison}), only our new measurements were considered.

\begin{deluxetable}{ccccc}
\tabletypesize{\scriptsize}
\tablecolumns{5}
\tablecaption{Additional H\,\scriptsize II \small regions from the literature\label{table:literature}}

\tablehead{
\colhead{\phantom{}ID\phantom{}}	     &
\colhead{\phantom{}R.A.\phantom{}}	 &
\colhead{\phantom{aaa}Decl.\phantom{aaa}}  &
\colhead{\phantom{aaa}R\phantom{aaa}}	 &
\colhead{\phantom{aa}other ID\phantom{aa}}	 \\[0.5mm]
\colhead{}       &
\colhead{(J2000.0)}   &
\colhead{(J2000.0)}   &
\colhead{(kpc)}   &
\colhead{} \\[1mm]
\colhead{(1)}	&
\colhead{(2)}	&
\colhead{(3)}	&
\colhead{(4)}	&
\colhead{(5)}	}
\startdata
\\[-2mm]
A &  	12~18~53.062	&	47~22~51.23	&  14.48	&	\zkh~9\\
B &  	12~18~26.791	&	47~21~09.23	&  23.30	&	\zkh~1, 5NA, B1   \\ 
   &					&				&		&	 OK~($-318 +183$)\\ 
C &  	12~18~57.445	&	47~17~35.83	&  2.51	&	\zkh~6 \\ 
&&&& OK~($-014 -038$)\\ 
D &  	12~18~43.814	&	47~17~32.21	&  16.04	&	\zkh~2, B2 \\
&&&& OK~($-152 -037$)\\ 
E &  	12~19~02.768	&	47~17~08.26	&   3.30	&	\zkh~5\\
F &  		12~18~55.470	&	47~16~47.82	&   7.27	&	OK~($-035 -086$) \\ 
G &  	12~18~59.308	&	47~16~44.46	&   4.47	&	\zkh~7\\
H &  	12~18~57.856	&	47~16~07.69	&   7.84	&	54C\\ 
I &  		12~18~58.614	&	47~15~56.09	&   8.00	&	58C\\ 
J &  		12~18~58.997	&	47~15~50.89	&   8.06	&	59C \\ 
&&&& OK~($-005 -150$) \\ 
K &  	12~19~00.116	&	47~15~36.95	&   8.17	&	69C\\
L &  		12~19~01.438	&	47~15~25.66	&   8.07	&	OK~($+024 -170$), 74C  
\enddata
\tablecomments{Units of right ascension are hours, minutes and seconds, and units of declination
are degrees, arcminutes and arcseconds. Col.~(1): \hii\/ region identification. Col.~(2): Right
Ascension. Col.~(3): Declination. Col.~(4): Galactocentric distance in kpc.
Col.~(5): Identification from \citet[=\,OK]{Oey:1993}, \citet[=\,\zkh]{Zaritsky:1994}, \citet[=\,B]{Bresolin:1999} and \citet[studied by \citealt{Diaz:2000}]{Courtes:1993}.\\}
\end{deluxetable}

\begin{figure*}
\medskip
\center \includegraphics[width=0.8\textwidth]{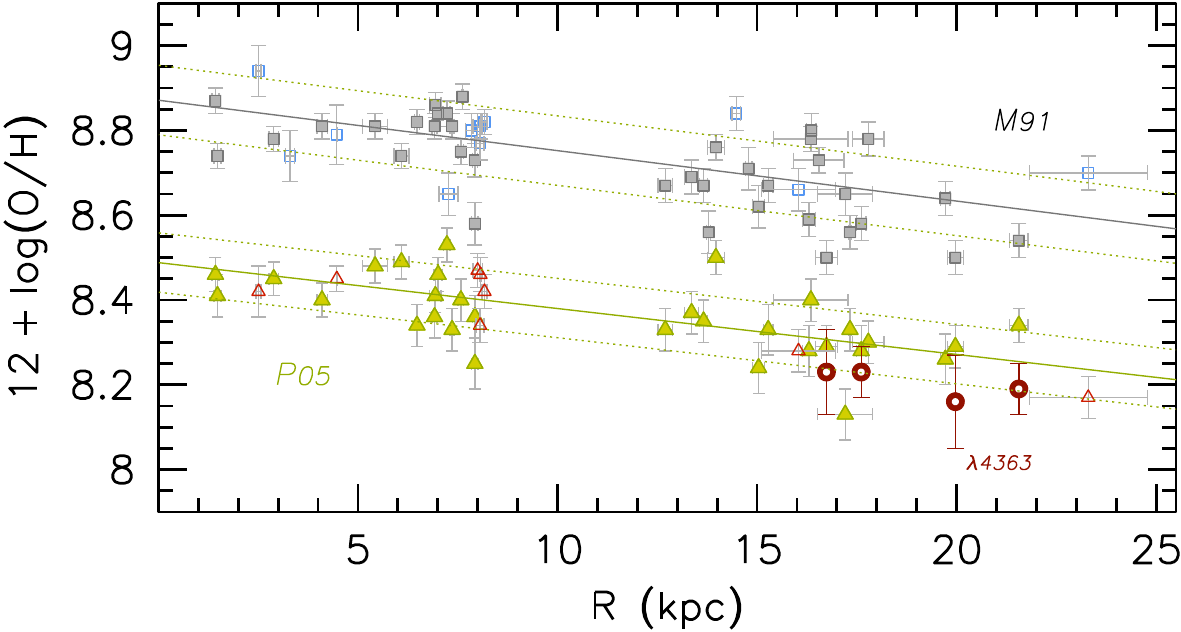}\medskip
\caption{Galactocentric oxygen abundance gradient in \ngc.  The abundances for the Gemini \hii\ region sample calculated with the M91 and P05 calibrations are shown as full squares and full triangles, respectively. Open symbols (squares and triangles for M91 and P05, respectively) are used for the \hii\ regions extracted from the literature. The corresponding regression lines, calculated accounting for errors in both coordinates, are shown by the grey and green lines.
The open circles show the oxygen abundance of the four \hii\ regions where the \oiii\lin4363 was detected. In several cases the errors in the deprojected galactocentric distances are smaller then the size of the symbols used.\\ \\
\label{fig:gradient}}
\end{figure*}

The galactocentric O/H gradient in \ngc\ is presented in Fig.~\ref{fig:gradient}. In this plot abundances for the Gemini \hii\ region sample calculated with the M91 and P05 \rtwothree\ calibrations are shown as full squares and full triangles, respectively. Open symbols (squares and triangles for M91 and P05, respectively) are used for the \hii\ regions extracted from the literature.
In the case of the abundances based on the P05 method objects with values of the excitation $P<0.3$ were removed, since the calibration provided by P05 only covers P\,$>$\,0.3, and we found that for objects with smaller excitation values the abundances could be overestimated.

For the calculation of the radial abundance gradient  a least-squares fit that accounts for errors in both coordinates (abundance and radius) was carried out. Because of the high inclination of the disk of \ngc\ relative to the line of sight (78$^\circ$), even a relatively small uncertainty in this angle can result into an appreciable error in the computed deprojected  
distances, especially for objects located near the minor axis. The source of the geometric parameters of the disk of \ngc\ adopted here, \citet{van-Albada:1980}, does not explicitly provide 
uncertainty estimates, but implies that the errors in the radio-determined position angle of the major axis and inclination are much smaller than 2$^\circ$ and 1$^\circ$, respectively. A 1$^\circ$ uncertainty was assumed for both angles. We also accounted for an astrometric uncertainty of 0\farcs4 in the \hii\ region positions.  The resulting error in the galactocentric distance reaches a maximum value of 6\%, but
for the majority  of the targets it is on the order of 1-2\%.

With the oxygen abundances  derived from the M91 calibration the regression to the data (grey line in Fig.~\ref{fig:gradient}) yields:

\begin{equation}
\rm \oh = 8.87~(\pm0.02)~-~0.012~ (\pm0.002)~R_{kpc}  \label{grad1}
\end{equation}

\noindent where $\rm R_{kpc}$ is the galactocentric distance measured in kpc. 
With the adopted values for the distance (7.2~Mpc) and the isophotal radius  \rtf\ (9\farcm31), the gradient slope  can also be expressed as  $-0.23\pm0.04$ dex\,\rtf\expone\ or $-0.025\pm0.004$ dex\,arcmin\expone.

In the case of the abundances derived from the {\em P} method of P05 the linear fit is (green line in Fig.~\ref{fig:gradient}):

\begin{equation}
\rm \oh = 8.49~(\pm0.02)~-~0.011~ (\pm0.002)~R_{kpc}    \label{grad2}
\end{equation}

\noindent
corresponding to $-0.21\pm0.03$ dex\,\rtf\expone\ or $-0.023\pm0.003$ dex\,arcmin\expone.
The dotted lines in Fig.~\ref{fig:gradient} show the 1\,$\sigma$ scatter around these fits (0.08 and 0.07 dex, respectively).

\bigskip
\noindent
It is worth pointing out the following:\\

\noindent
- the slopes of the two fits agree within the uncertainties, but the intercepts  differ by 0.38 dex. This is due to the aforementioned systematic offsets between different calibrations of strong-line indicators, in particular between one calibrated from photoionization models (M91) and one calibrated empirically (P05).  \\[-2mm]

\noindent
- a look at Fig.~\ref{fig:gradient} suggests that 
in the inner 8 kpc the gradient is virtually flat, with constant \oh\,=\,$8.79\pm0.07$ (M91)  and \oh\,=\,$8.40\pm0.07$ (P05), however this solution is not found to be statistically preferable over the single slope solution.   \\[-2mm]

\noindent
- the result on the abundance gradient {\em slope} appears to be relatively robust. We considered the \rtwothree\ calibrations by \citet{Zaritsky:1994} and \citet{Tremonti:2004}, both of which are based on theoretical models, and  obtained results that are in agreement with the two previous determinations, as Table~\ref{table:regression} shows. It should also be noted that the \citet{Zaritsky:1994} calibration yields the largest intercept value, 0.53~dex above the P05 calibration and 0.15~dex above the M91 calibration.
\\[2mm]

\begin{deluxetable*}{lcccc}
\tabletypesize{\scriptsize}
\tablecolumns{5}
\tablecaption{Results of linear regression to the abundance gradient\label{table:regression}}
\tablehead{
\colhead{\phantom{aaaaaa}\rtwothree\ calibration\phantom{aaaaaa}}	 &
\colhead{Intercept}  &
\multicolumn{3}{c}{Slope} \\[0.5mm]
\colhead{}       &
\colhead{}   &
\colhead{dex\,kpc\expone}   &
\colhead{dex\,\rtf\expone}   &
\colhead{dex\,arcmin\expone}   \\[1 mm]
\colhead{(1)}	&
\colhead{(2)}	&
\colhead{(3)}	&
\colhead{(4)}	&
\colhead{(5)}	}
\startdata
\\[-2mm]
\multicolumn{5}{c}{\em This paper}\\[2mm]
\citet{McGaugh:1991}		&	$8.87 \pm 0.02$   & $-0.012 \pm 0.002$  		&	$-0.23\pm0.04$	&	$-0.025\pm0.004$	\\  
\citet{Pilyugin:2005a}			&	$8.49 \pm 0.02$   & $-0.011 \pm 0.002$  		&	$-0.21\pm0.03$	&	$-0.023\pm0.003$	\\ 
\citet{Zaritsky:1994}			&	$9.02 \pm 0.03$   & $-0.013 \pm 0.002$  		&	$-0.25\pm0.05$	&	$-0.027\pm0.005$	\\
\citet{Tremonti:2004}			&	$8.94 \pm 0.02$   & $-0.011 \pm 0.002$  		&	$-0.22\pm0.04$	&	$-0.023\pm0.005$	\\[3mm]
\multicolumn{5}{c}{\em Published}\\[1mm]
{\em Authors}	&	   &  		&		&		\\[1mm]
\citet{Zaritsky:1994}		&	$9.17 \pm 0.06$   & $-0.029 \pm 0.005$  		&	$-0.58\pm0.09$	&	$-0.062\pm0.010$	\\
\citet{Dutil:1999}		&	$8.86 \pm 0.02$   & $-0.012 \pm 0.002$  		&	$-0.23\pm0.04$	&	$-0.025\pm0.004$	\\
\citet{Pilyugin:2004}		&	$8.57$ 		   & $-0.010$ 		  		&	$-0.20$	&	$-0.021$					\\
\citet{Bono:2008}		&	$8.55$ 		   & $-0.010$ 		  		&	$-0.20$	&	$-0.021$	
\enddata
\end{deluxetable*}

The data points corresponding to the \oiii\lin4363-based abundances are shown as red circles (labeled \lin4363) in Fig.~\ref{fig:gradient}. Their positions in the plot are within $\sim$1\,$\sigma$ of the regression line for the abundances of the full sample based on the P05 method. This relatively good agreement is not surprising, since the {\em P} method was calibrated empirically from \hii\ regions whose abundances were obtained from the use of the auroral lines. For Cepheid-related work it is worth pointing out that the abundances tied to the \oiii\lin4363 detections (i.e.~those obtained from the P05 calibration) are to be considered in the \te-based scale (\citealt{Kennicutt:2003}, \citealt{Sakai:2004}).
Although the overall metallicity of \ngc\ in the \te\ scale might seem to be low, its characteristic oxygen abundance [\oh\,=\,8.40 at $R$\,=\,0.4\,\rtf] is normal for this galaxy's luminosity and rotational velocity, according to the trends found
by \citet{Pilyugin:2004} using abundances of galaxies obtained from the $P$ method, albeit with an O/H ratio slightly below average.
We can also compare \ngc\ to a galaxy with a well-determined, \te-based abundance gradient like M101, which has an absolute luminosity 
$M_B = -20.9$, similar to that of \ngc\ ($M_B = -20.8$). The slope of the abundance gradient in M101 is $-0.032$~dex\,kpc\expone,
with an extrapolated central abundance \oh\,=\,$8.75 \pm 0.05$ (\citealt{Bresolin:2007}). The $\sim0.26$~dex lower intercept for \ngc\ can be understood by recalling  that strongly barred galaxies\footnote{Evidence for the presence of a bar in \ngc\ has been provided by \citet{van-Albada:1980}.}, in addition to shallower gradients, also display significantly lower (0.3-0.4 dex) central abundances compared to non-barred galaxies (\citealt{Dutil:1999}; see also Fig.~11 of \citealt{Vila-Costas:1992}).

As already pointed out, the absolute values of the chemical abundances are greatly dependent on the
choice of abundance determination method. The {\em P} method by P05 provides results which are in good agreement with the auroral line-based abundances determined for a handful of objects. However, comparisons with independent methods would be  highly desirable, in order to establish the absolute abundance scale in \ngc. This is relevant, for example, for the amount of the metallicity correction to the Cepheid-based distance modulus.
Future spectroscopic studies of the stellar abundances in this galaxy are already planned by the author and his collaborators with observations at the Keck telescope. It is also interesting to note that the recent work by \citet{Konami:2009} obtained a metallicity of the interstellar medium in \ngc\ from observations with the Suzaku X-ray satellite. Their result, $\sim$0.5\,\zsun\ in the \citet{Anders:1989} solar metallicity scale, corresponds approximately to \oh\,=\,8.6, i.e.~an intermediate value between those indicated by the two methods used here.\\


\section{Previous determinations of the abundance gradient}\label{section:others}
The oxygen abundance gradient of \ngc\ was already derived as part of earlier spectroscopic studies of large samples of \hii\ regions  in different galaxies, in particular by \citet[=\,\zkh]{Zaritsky:1994}, who also included spectrophotometric data obtained by \citet{Oey:1993}, for a total number of  about 15 objects. However, 
before a direct comparison with our results can be made it is necessary to homogenize the distances and diameters adopted by the different authors. In particular, \zkh\ adopted an isophotal radius of 7\farcm92 from the RC2 (\citealt{de-Vaucouleurs:1976}), which differs significantly from the value of 9\farcm31 that we adopted from the RC3 (\citealt{de-Vaucouleurs:1991}). Thus, the gradient published by \zkh:

$$\rm \oh = 8.97~ (\pm 0.06)~-~ 0.49~ (\pm0.08)~(R/R_{25}- 0.4)$$

\noindent
becomes, after renormalization:

$$\rm \oh = 9.166 ~ (\pm 0.06)~ -~ 0.029~ (\pm 0.005) ~ R_{kpc}$$

\noindent
This is a significantly steeper gradient than the one we have derived earlier in Eq.~(\ref{grad1}) and ~(\ref{grad2}). It appears that the \zkh\ result is highly dependent on the abundance calculated  for one of the two outermost \hii\ regions, number 8 in their list (our number 30, for which we also obtained a \oiii\lin4363 detection). The problem is not in the line fluxes, since we measure very similar \oii\lin3727 and \oiii\llin4959,\,5007 line intensities.
However, the application of \zkh's own calibration of \rtwothree\ provides  quite a low oxygen abundance for this particular \hii\ region, \oh\,=\,8.41,  about 0.4 dex lower than other objects at a similar galactocentric distance. This anomaly appears to be related to the high \rtwothree\ value. In fact, two additional \hii\ regions in our sample 
(24 and 36), which also have log(\rtwothree)\gt0.9, display peculiarly low O/H abundances when adopting the \zkh\ and the \citet{Tremonti:2004} \rtwothree\  calibrations. This suggests  a systematic problem with the functional form of these calibrations at high \rtwothree\ values.
Therefore, we have excluded these three \hii\ regions to calculate the regressions whose parameters are shown in Table~\ref{table:regression}.

\medskip
As part of an abundance study of eight early-type spiral galaxies, \citet{Dutil:1999} measured the abundance gradient of  \ngc, using a  spectrophotometric technique based on narrow-band imaging through filters centered on important nebular emission lines and applied to 122 \hii\ regions. They adopted as their abundance indicator the \oiii\lin5007/\nii\lin6583 line ratio, calibrated as a function of O/H by \citet{Edmunds:1984}. These authors adopted a distance to \ngc\ of 7.3~Mpc, i.e.~virtually the same one used here. They obtained the following oxygen abundance gradient:

$$\rm \oh = 8.86 ~ (\pm 0.02)~ -~ 0.012~ (\pm 0.002) ~ R_{kpc}$$

\noindent 
which is the same we derive from our sample using the M91 \rtwothree\ calibration. \citet{Dutil:1999} do not provide a list of line fluxes for their targets, so that we cannot compare their measurements with ours on an object-by-object basis.  However, the excellent agreement between the two gradient determinations
indicates that the imaging spectrophotometry of \citet{Dutil:1999} is not affected by important systematic uncertainties, which could originate especially from the subtraction of the stellar underlying continuum, and which could introduce biases in the chemical abundances derived with this technique (\citealt{Dutil:2001}).

\medskip
More recently, \citet{Bono:2008} reconsidered the abundance gradient in \ngc\ in their reanalysis of the Cepheid variables discovered by \citet{Macri:2006}. \citet{Bono:2008} found that the comparison of the data with pulsation models 
suggested both a lower metallicity and a shallower abundance gradient in \ngc\ than one obtains from the \zkh\ paper. Taking the few data points published by \citet{Diaz:2000}, these authors derived an abundance gradient which, after renormalization to the quantities adopted in our paper,  has a slope of $-0.010$ dex\,kpc\expone, with a central abundance \oh\,=\,8.55. This agrees also with the study by \citet{Pilyugin:2004}, who used the {\em P} parameter method to measure abundance gradients from data in the literature for more than 50 galaxies.
Finally, \citet{Riess:2009a} obtained abundances from Keck observations. They do not provide information about their targets and their derived abundances, but from their Table~12 we infer a slope of $-0.017$ dex\,kpc\expone. This gradient is possibly too steep because of the presence of two \hii\ regions at large galactocentric distance and low O/H ratio (as judged from their Fig.~11), which might suffer from the same large-\rtwothree\ issue mentioned earlier.
\medskip

To summarize, the evidence from most of the results published in the literature and from the analysis of the new Gemini data presented here is 
that the abundance gradient in \ngc\ is very shallow, with a slope of approximately $-0.012$ dex\,kpc\expone, in contrast with the steeper (by a factor of 2.5) slope obtained by \zkh. This shallow abundance gradient is not surprising, in view of the fact that barred galaxies (like \ngc) and spiral galaxies of early type 
have smaller slopes compared to non-barred and late-type galaxies (e.g.~\citealt{Vila-Costas:1992}, \citealt{Zaritsky:1994}).
For example, \citet{Vila-Costas:1992} found that the barred galaxies in their sample have slopes that are shallower than $-0.05$ dex\,kpc\expone, while non-barred galaxies can have slopes up to four times steeper. This behavior is also predicted by numerical simulations of barred galaxies
(\citealt{Friedli:1994}).

\begin{figure}
\medskip
\epsscale{1.15}
\plotone{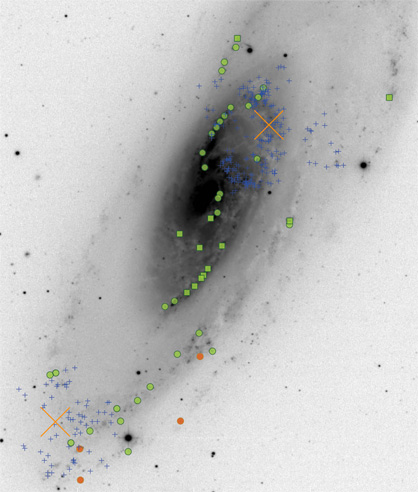}
\caption{The distribution of Cepheids in \ngc\ (\citealt[blue crosses]{Macri:2006}) compared with the location of the \hii\ regions studied in this work (circles for the new Gemini sample, squares for the objects drawn from the literature). The red circles represent the \hii\ regions with the \oiii\lin4363 detections. The big crosses show the centers of the two ACS fields observed by \citet{Macri:2006}.
$B$-band image courtesy of L.~van Zee.\\
\label{fig:cepheids}}
\epsscale{1}
\end{figure}

\section{The metallicity dependence of the Cepheid Period-Luminosity relation in \ngc}

\citet{Macri:2006} have identified  281 Cepheid variables in two {\em HST} ACS fields of \ngc, whose centers are located 179$''$ (6.25~kpc, `inner' field) and 490$''$ (17.11~kpc, `outer' field) from the galaxy nucleus (deprojected distances). Our spectroscopic coverage of \hii\ regions in \ngc\ was designed to allow the measurement of  nebular abundances at the same galactocentric distances of the Cepheids, as visualized in Fig.~\ref{fig:cepheids}.
The extinction-corrected distance moduli of the two fields relative to the LMC, as determined by \citet{Macri:2006} from the {\em BVI} \perlum\ relations of a restricted sample of Cepheids, are $\Delta\mu_o = 10.87$ (inner field) and $\Delta\mu_o = 10.71$. These authors ascribed this result to the difference 
in metallicity between the two fields, as a consequence of the dependence of the \perlum\ relation on metal content. Adopting the radial abundance gradient measured by \zkh, they fitted the distribution of the distance moduli derived for individual Cepheids as a function of the O/H ratio, to obtain a value of the parameter $\gamma = \delta\mu_0/\delta\log Z = -0.29\pm0.09~\rm (random) \pm 0.05~ (systematic)$ mag\,dex\expone. 
This $\gamma$ value is in good agreement with other empirical determinations (see a review of the extensive data present in the literature by \citealt{Romaniello:2008}).
Here, we just mention some of the recent results based on the differential analysis of Cepheids located at different 
distances from the center of their host galaxy. \citet{Kennicutt:1998} determined a value of $\gamma = -0.24\pm0.16$ mag\,dex\expone\ from {\em HST} observations  of two fields in M101. More recently, \citet{Scowcroft:2009} performed a similar experiment in the galaxy M33, obtaining $\gamma = -0.29\pm0.11$ mag\,dex\expone. It is well-known that these empirical results, showing that metal-rich Cepheids are brighter than metal-poor ones, are at variance with the predictions from pulsation models, i.e.~that Cepheids become {\em fainter} with increasing metallicity, with a dependence on the passbands of the observations (\citealt{Fiorentino:2007}). The latter finding is backed by recent investigations of the iron content of Galactic and Magellanic Cloud Cepheids (\citealt{Romaniello:2008}).
\footnote{The $\gamma$ values reported here refer to distances measured from {\em VI} photometry. The metallicity effect is predicted to vary with the use of different passbands.}

For the case of \ngc, as demonstrated in Sect.~\ref{section:oxygen} and \ref{section:others}, the \zkh\ abundance gradient adopted by \citet{Macri:2006} is much steeper than our new observations and other data in the literature indicate. Instead of a difference in oxygen  abundance of $\rm\delta\log(O/H) = 0.31$~dex between the inner and outer fields, as one would derive from the \zkh\ gradient, we measure $\rm\delta\log(O/H) \simeq 0.13$~dex.
Correspondingly the metallicity correction factor  would become $\gamma = -0.69$ mag\,dex\expone, an unacceptably large value from both the empirical and theoretical standpoints. We are drawn to  conclude, together with \citet{Bono:2008}, that, because of its shallow gradient, \ngc\ (like any other barred galaxy)  is not the ideal laboratory to perform  differential studies of the metallicity dependence of the Cepheid pulsation properties.

What is then the origin of the observed distance modulus difference between the inner and outer fields in \ngc?
While we cannot presently discard the effects of varying Cepheid helium abundance or differences in the reddening law (\citealt{Mager:2008}), the presence of systematic errors affecting the Cepheid distance of the outer field appears as a possible solution  (\citealt{di-Benedetto:2008}). In particular, \citet{Mager:2008} showed how the outer field Cepheids, contrary to the inner field ones, do not evenly populate the instability strip, projected into the \perlum\ relation. Instead, their magnitudes are skewed towards faint levels, thus artificially increasing the distance modulus.

\subsection{The metallicity of \ngc\ on the \te\ scale}\label{sec:tescale}
The detection of the \oiii\lin4363 line and the derivation of O/H ratios from the strength of this line in four of the outer \hii\ regions has allowed us to place the chemical abundance of \ngc\ on the \te\ scale on rather solid evidence, in combination with  the use of the P05 calibration of the \rtwothree\ metallicity indicator for the innermost part of the galaxy. From Eq.~\ref{grad2} we obtain:

\begin{trivlist}
\item {\em inner field:} \oh\eq $8.42\pm0.03$

\item {\em outer field:} \oh\eq $8.30\pm0.06$
\end{trivlist}

If for the outer field we take the \lin4363-based abundance for objects 24 and 34 (their galactocentric distance matches that of the center of the outer ACS field) we obtain 12 + log(O/H) = $8.23\pm0.09$. Given the size of the errors, and the statistical nature of the \rtwothree\ indicator, we do not regard this small difference as a significant discrepancy, and adopt the former result for consistency with the inner field abundance, which can only be derived from the linear regression.
\medskip

\begin{figure}
\medskip
\epsscale{1.15}
\plotone{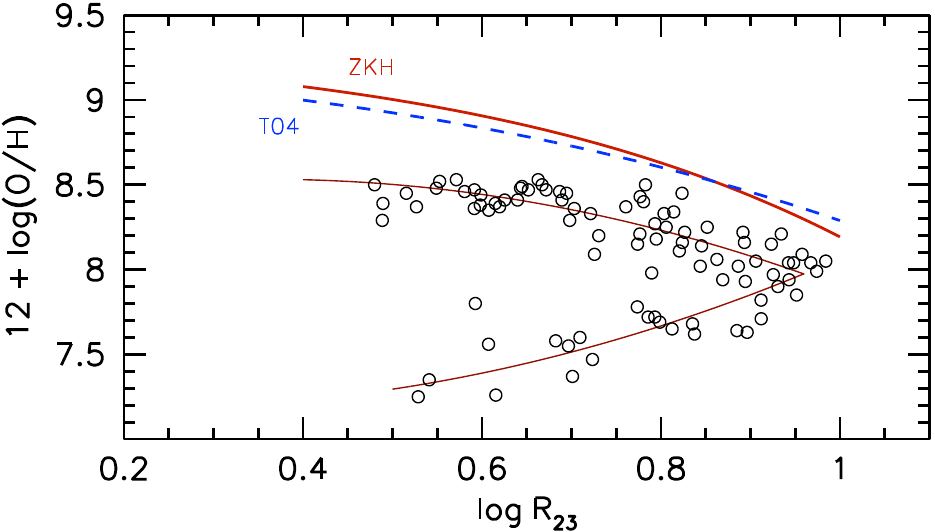}
\caption{Log(\rtwothree) \vs\ \oh\ for the same sample of extragalactic \hii\ regions shown in Fig.~\ref{fig:r23}. The curves in the upper part represent the \rtwothree\ calibrations by \citet[continuous line, \zkh]{Zaritsky:1994} and \citet[dashed line, T04]{Tremonti:2004}.\\
\label{fig:r23_2}}
\end{figure}

It is interesting to compare these abundances with those proposed by other authors working on the Cepheid distance scale. \citet{Sakai:2004} obtained a transformation
between nebular oxygen abundances of Cepheid host galaxies based on the \zkh\ abundance scale, and those one derives using the auroral line method, which is based on the ability to directly measure the electron temperature of the ionized gas (hence the `\te-based' definition). As already stressed, the \te\ abundances are systematically lower than those derived from the \zkh\ and most of the other strong-line metallicity indicators (\citealt{Kennicutt:2003, Bresolin:2004}). The effect is stronger at high metallicity, thus compressing the metallicity range spanned by star-forming galaxies.
This can be appreciated in Fig.~\ref{fig:r23_2}, where the same data points of Fig~\ref{fig:r23} are shown, together with the curves corresponding to the \zkh\ (continuous line) and \citet[dashed line]{Tremonti:2004} \rtwothree\ calibrations. These two curves are seen to be diverging from the 
\te-based abundances, shown by the dots, as the metallicity increases, with the effect being stronger in this example for the \zkh\ calibration.

It should also be recalled that most of the nebular work of the past few years has adopted the recent solar oxygen abundance derived from 
3-D hydrodynamical models of the Sun, which yield \oh$_\odot$\,=\,8.69 (\citealt{Allende-Prieto:2001, Asplund:2009}), 0.24 dex below the \citet{Anders:1989} value.
In this scale, for example, the LMC is found to have a nebular oxygen abundance of \oh$_{\rm LMC}$\,=\,$8.36\pm0.10$. This value is the mean obtained by the author (see Appendix) from a reanalysis of published emission line fluxes carried out with consistent atomic data.
This is in excellent agreement with results from other stellar indicators, for example Cepheids, for which 
 \citet{Romaniello:2008} found [Fe/H]$\rm_{LMC}$\,=\,$-0.33\pm0.13$, corresponding to \oh\,=\,$8.36\pm0.13$.
 
\medskip

\begin{figure}
\medskip
\epsscale{1.15}
\plotone{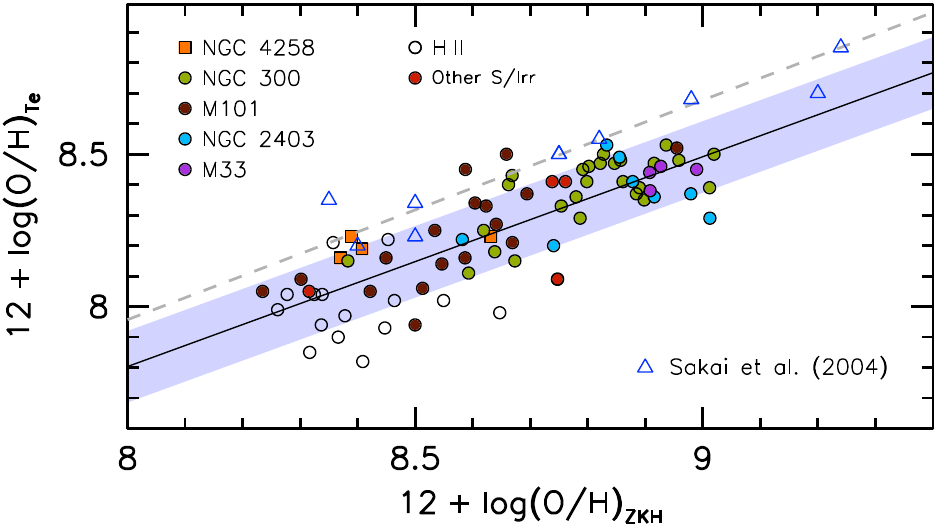}
\caption{Comparison between the oxygen abundances in the \zkh\ and \te\ (i.e.~\oiii\lin4363-based) scales. The sample of \hii\ regions is the same as in Fig.~\ref{fig:r23}. The linear regression to the \hii\ region sample is shown by the straight line, and the $\pm 1\sigma$ scatter is represented by the shaded area. The high-metallicity (\rtwothree\ upper branch) galaxy data points of \citet{Sakai:2004} are shown by the blue triangles. The dashed line is the linear fit to the latter by \citet{Saha:2006}.\\ 
\label{fig:zkh}}
\end{figure}

\citet{Saha:2006} and \citet{Bono:2010} performed polynomial fits to the \citet{Sakai:2004} data points, producing a relation between O/H abundance ratios in the \zkh\ system and the \te\ system. Using the data in Table~7 of \citet{Bono:2010} we find that in the case of \ngc\ the transformed \oh\ abundances are 0.2 dex larger than  the abundances that we measure: 8.64 \vs\ 8.42 for the inner field, 8.50 \vs\ 8.30 for the outer field. 
To investigate the origin of this discrepancy,  the relation between the oxygen abundances in the \zkh\ and \te\ scales for the same sample  of \hii\ regions shown in Fig.~\ref{fig:r23} was considered. The result is illustrated in Fig.~\ref{fig:zkh}, where the abundances from Table~4 of \citet[open triangles]{Sakai:2004} and the linear fit to the latter by \citet[dashed line]{Saha:2006} are included. Only points in the \rtwothree\ upper branch are shown, because the \zkh\ calibration only applies to that regime (the transformation between the two abundance scales below \oh$_{\rm ZKH}$\,$\simeq$\,8.2 is thus undefined).
The linear fit to the individual \hii\ region data (continuous line) in Fig.~\ref{fig:zkh}  lies 0.15 dex below the \citet{Saha:2006} regression at 
$x$\,=\,\oh$_{\rm ZKH}$\,=\,8.5.
The new transformation is:

\begin{equation}
\rm \oh_{Te} = 0.69~(\pm 0.06)~ {\mathit x}~ +~ 2.30~(\pm 0.51).\label{eq:zkh}
\end{equation}

\noindent
Only two data points from \citet{Sakai:2004}, the inner and outer fields of M101, fall within the 1\,$\sigma$ scatter ($\pm 0.11$\,dex) of this regression. The origin of the \te-based abundances
provided by  \citet[Table~4]{Sakai:2004} is not specified in their paper, so it is difficult to ascertain the reason for the discrepancy for the remaining upper branch metallicities. It should also be added that at the time of their publication the systematic offset between strong-line abundances and direct abundances in \hii\ regions had just become clear (\citealt{Kennicutt:2003}), and that since then the number of high-metallicity \hii\ regions with auroral line detections has increased significantly (e.g.~\citealt{Bresolin:2009a}).

The comparison between our new and the previous \zkh-\te\ abundance transformations explains the reason why the metallicities we derive for the two fields in \ngc\ are smaller than those obtained from the 
\citet{Saha:2006} or \citet{Bono:2010} transformations.
It also indicates that the latter should be revisited according to Eq.~\ref{eq:zkh} if one wishes to transform in a consistent way the strong-line abundances in the \zkh\ scale to the \te\ scale. This is done in the Appendix for a number of Cepheid host galaxies.
\medskip

We would also like to point out that the fact that the multiplicative  coefficient for $x$ in Eq.~\ref{eq:zkh} differs significantly from unity (the same information is conveyed by the 
divergence of the \zkh\ curve in Fig.~\ref{fig:r23_2} from the \te-based empirical abundances with increasing metallicity)
 implies that abundance gradient slopes can depend on the choice of abundance determination methods. The effect is small in the case of \ngc, where the gradient slope is very shallow and the metallicity range over the full extent of the galactic disk is $\sim0.2$ dex. However, for galaxies with steeper abundance gradients the effect could become significant.

\subsection{Consequences for the Cepheid distances}
The abundances in the \te\ scale place the inner field of \ngc\ essentially at the metallicity of the LMC, rather than at the Galactic value (\oh\,$\simeq$\,8.7), which has been usually adopted in the literature to estimate the distance modulus to \ngc\ (e.g.~\citealt{Benedict:2007}). How would this affect the Cepheid distance to \ngc\ from the \citet{Macri:2006} data? To answer this question, two methods have been considered:\\

\noindent
{\em a)} we take the extinction-corrected distance modulus of the inner field of \ngc\  relative to the LMC from \citet{Macri:2006}, $\Delta\mu_o = 10.71 \pm 0.04_r \pm 0.05_s$, and apply no metallicity correction, since the O/H abundance is the same. Thus, 
adopting $\mu_{\rm LMC}$\,=\,18.5 from the {\em HST} Key Project (\citealt{Freedman:2001}), we obtain $\mu_{\rm NGC~4258}$\,=\,$29.21\pm 0.04_r \pm 0.05_s$. Alternatively, if we assume the maser distance modulus to \ngc, $29.29 \pm 0.09_r \pm 0.12_s$ we would obtain  $\mu_{\rm LMC} = 18.58 \pm 0.10_r \pm 0.13_s$.  \\

\noindent
{\em b)} we follow \citet{Mager:2008}, who restricted the \citet{Macri:2006} sample to Cepheids with periods above 10 days, for consistency with the procedure adopted by the Key Project. For the inner field they obtain $\mu_0 = 29.26 \pm 0.03$. The metallicity correction applied by the Key Project is $\rm \delta\mu_z = -0.2[ (O/H) - (O/H)_{LMC}]$ mag\,dex\expone. Since we are testing the idea that the inner field has the same metallicity of the LMC, this correction would be zero in our case. Thus, $\mu_{0, NGC~4248} = 29.26 \pm 0.03$.\\

\indent
We also note that the adoption of an LMC-like metallicity for the inner field explains the finding by \citet{Tammann:2008} that the {\em BVI} \perlum\ relations in this field agree with those in the LMC, rather than those for the higher Galactic metallicity, without the need to invoke the 
existence of a second parameter, such as the helium abundance. In virtue of the shallow abundance gradient, the outer field is also at a similar metallicity.
We can also conclude that the adoption of the \te\ abundance scale for \ngc, as presented here, would yield results that are compatible with the Key Project and with the geometric distance from the water masers.

\subsection{Reliability of the absolute  \te\ abundances}
Although our analysis of \hii\ region data can be used to place the metallicity of the interstellar medium of \ngc\ on the \te\ scale, there are some important considerations that need to be taken into account when dealing with nebular abundances obtained from the \oiii\lin4363 auroral line.
First is the effect of dust depletion onto dust grains. Recent estimates of the oxygen depletion factor obtained from the study of the Orion nebula are in the range between 0.08~dex (\citealt{Esteban:2004}) and 0.12 dex (\citealt{Mesa-Delgado:2009}). Secondly, accounting for the presence of temperature fluctuations in the ionized gas has the effect of increasing the metal abundances (\citealt{Peimbert:1967}),
yielding results that are compatible with those from metal recombination lines.
For these reasons, the \te-based abundances are often regarded as lower limits to the real abundances.

In the case of the Orion nebula, whose \te-based oxygen abundance is 12 + log(O/H) = $8.51\pm0.03$ (\citealt{Esteban:2004}),
the sum of these two effects is about 0.25 dex. Recently \citet{Simon-Diaz:2010} have shown 
that the oxygen abundance for the gas phase (from metal recombination lines) in the Orion nebula, increased by the amount locked in dust grains,  is consistent with the value of \oh\,=\,$8.74\pm0.04$ derived from 13 B stars in the Ori OB1 association. This compares very well with the Galactic neighborhood value of \oh\,=\,$8.76\pm0.03$, calculated by \citet{Przybilla:2008} from B stars. 
These results suggest that the \te-based results could systematically underestimate the oxygen abundance by 0.25 dex. 

On the other hand, the addition of the depletion correction factor to the recombination line abundance of 30~Dor in the LMC gives 
an abundance, \oh\,=\,$8.61\pm0.03$ (\citealt{Peimbert:2010}), which is clearly at variance with respect to the B-star
[\oh\eq$8.33\pm0.08$, \citealt{Hunter:2007}]
 and Cepheid (\citealt{Romaniello:2008}) metallicities, unless it is assumed that the present-day chemical composition of the LMC is highly non-uniform.
Therefore, in this case the \te-based oxygen abundance is in excellent agreement with the stellar one, as is also found for the SMC (e.g.~\citealt{Trundle:2005}), where \oh\eq $8.03\pm0.09$ (see Appendix).\bigskip\bigskip

\subsection{The case of M33}
We also briefly consider the case of M33, where  \citet{Scowcroft:2009} have determined $\gamma = -0.29\pm0.11$ mag\,dex\expone\ from the
difference in distance modulus between an inner and an outer field, and adopting a metallicity gradient obtained from \hii\ regions
which is very steep ($-0.19$\,dex\,kpc\expone) in the central 3~kpc, and rather shallow ($-0.038$\,dex\,kpc\expone) beyond this radius  (taken from \citealt{Magrini:2007}). The choice of this bimodal abundance gradient is questionable, and  the most recent works on the chemical composition of the ionized gas 
(\citealt{Crockett:2006, Rosolowsky:2008, Bresolin:2010})
and young supergiants (\citealt{U:2009}) in M33 yield a single-slope gradient, with a shallow value throughout the disk of the galaxy. The most recent and arguably one of the most accurate determinations to date of the \hii\ region radial oxygen abundance gradient in M33 has been obtained by \citet{Bresolin:2010}, who measured 
a slope of $-0.030\pm0.008$ dex\,kpc\expone\
from \oiii\lin4363 detections, which extend into the inner kpc of the galaxy. They found no indication of an upward turn in the nebular abundances in the inner 3~kpc, where they measured \oh\,$\simeq$\,8.4, instead of \oh\,=\,8.85, as  used by  \citet{Scowcroft:2009}.
The adoption of such a shallow abundance slope yields a total range in O/H between the inner and outer regions of M33 of only 0.14\,dex.
This, combined with the distance moduli obtained at the two different galactocentric distances by 
 \citet{Scowcroft:2009}, would result in an exceedingly strong metallicity effect on the Cepheid-derived distances, with $\gamma\simeq-1.2$ mag\,dex\expone\ .

We note that a bias from small number statistics, similar to the one encountered in \ngc, could also be affecting the M33 result by \citet{Scowcroft:2009}. As can be seen from the \perlum\ and Period-Wesenheit relations shown
in their Fig.~5 and 6, the scatter for the outer field Cepheids at constant period is much smaller than that for the inner field, indicating that the width of the instability strip is not sufficiently sampled. This effect is exacerbated by the scarcity of Cepheids with periods $\ge$ 10 days in their outer field (only two are found). 
We reach the same conclusion we drew for \ngc, that  $\gamma$ cannot be reliably derived in M33 with present Cepheid samples. A more reliable approach to constrain the metallicity effect on Cepheid-based extragalactic distances is represented by the comparison with the distances provided by the \trgb\ method as a function of the host galaxy metallicity (\citealt{Sakai:2004, Tammann:2008, Bono:2010}), provided the latter is measured consistently, for example in the \te\ scale, as explained in \S~\ref{sec:tescale}.

\section{Summary}

This paper combined new deep spectroscopic observations of 36 \hii\ regions in \ngc\ obtained with the Gemini telescope with data from the literature, to measure the oxygen abundance gradient for this galaxy. Strong-line abundances have been derived from two different calibrations of the \rtwothree\ indicator, the one by \citet{McGaugh:1991} and the one by \citet{Pilyugin:2005a}. For four of the outermost \hii\ regions the \oiii\lin4363 auroral line has been detected, and used to measure the nebular abundances directly in the \te\ scale. The main results are summarized here:

\begin{trivlist}

\item 1. The radial abundance gradient in \ngc\ has a slope of approximately $-0.012$ dex\,kpc\expone, which is a factor of 2.5 shallower than the value reported by \citet{Zaritsky:1994}
and adopted by \citet{Macri:2006} to infer a metallicity correction of Cepheid distances of $\gamma =  -0.29$ mag\,dex\expone.

\item 2. Adopting the shallow gradient resulting from the new analysis would yield an unrealistically large metallicity correction for the Cepheid distances. This indicates that \ngc, with its present sets of Cepheid observations, is not suited for a differential metallicity study to measure $\gamma$ reliably.

\item 3. The \oiii\lin4363 detections and the P05 \rtwothree\ calibration provide the oxygen abundances in the \te\ scale. In this scale, the inner field of \ngc\ has an LMC-like O/H ratio, and not a Galactic one.

\item 4. The same systematic biases from small number statistics that might explain the difference in distance moduli between the inner and outer regions of \ngc\  (\citealt{Mager:2008}) seem to affect also the differential analysis of M33 Cepheids by \citet{Scowcroft:2009}.

\item 5. The transformation between `old' \hii\ region metallicities in the \citet{Zaritsky:1994} system and the \te\ system by \citet{Sakai:2004} is affected by a systematic error of about 0.2 dex. Using recent observations of \hii\ regions with auroral line-based abundances, a revised transformation has been calculated. 

\bigskip

The main results of this work can be tested in the near future, with the availability of larger samples of Cepheids in the outer regions of \ngc\ and M33, to remove potential statistical biases, and with the derivation of blue supergiant metallicities, which will help to establish the absolute scale of the chemical abundances in \ngc.

\end{trivlist}

\acknowledgments
FB gratefully acknowledges the support from the National Science Foundation grants AST-0707911 and AST-1008798, and thanks R.P.~Kudritzki for discussions and comments on an earlier draft of this paper.
The Gemini observations used in  this paper were obtained under program GN-2010A-Q-25.

\medskip
\noindent
{\it Facility:} \facility{Gemini:Gillett (GMOS)}\\

\bibliography{ngc4258}


\appendix

\section{Metallicities of Cepheid host galaxies in the \te\ scale}
In this Appendix the oxygen abundances of Cepheid host galaxies are calculated self-consistently, using {\em (a)} observations of the \oiii\lin4363 auroral line
 or {\em (b)} the conversion given in Eq.~\ref{eq:zkh} between abundances obtained from the  \citet[\zkh]{Zaritsky:1994} calibration of \rtwothree\ and \te-based abundances. This equation was obtained by comparing upper branch abundances for a large sample of \hii\ regions measured using the two methods.
 It should be recalled that the \zkh\ calibration only holds for the metal-rich branch of \rtwothree, and therefore the transformation is not defined for the low-metallicity branch (\oh$_{Te}$~$<$~8).

\subsection{Irregular and dwarf galaxies with abundances determined from \oiiiit\lin4363 detections. }
Emission line data for nine irregular and dwarf galaxies with \oiii\lin4363 line detections were drawn from the literature, and the oxygen abundances for individual \hii\ regions were calculated  using the same method ({\sc iraf}'s {\tt\small temden} and {\tt\small ionic} routines) and atomic data (see \citealt{Bresolin:2009a} for details) used for our \te-based abundance determinations in \ngc.
The average abundances and standard deviations are reported in Table~\ref{table:irregulars}, together with the literature sources for the emission line data.
\bigskip

\begin{deluxetable}{lcl}
\tabletypesize{\small}
\tablecolumns{3}
\tablewidth{0pt}
\tablecaption{Cepheid host irregular galaxies with \oiii\lin4363 detections \label{table:irregulars}}

\tablehead{
\colhead{\phantom{}Galaxy\phantom{aaaaaaaaaa}}	     &
\colhead{\phantom{aaa|}\oh\phantom{aaa}}       &
\colhead{Sources of emission line data}  }
\startdata
\\[-2mm]
LMC\dotfill		& 	8.36 $\pm$ 0.10			&		 \citet{Dufour:1975, Pagel:1978, Dufour:1982}					\\
				&									&		 \citet{Russell:1990,Tsamis:2003}							\\[0.5mm]
SMC\dotfill		& 	8.03 $\pm$ 0.09			&		 \citet{Dufour:1975, Dufour:1977, Pagel:1978}					\\
				&									&		 \citet{Russell:1990,Tsamis:2003}							\\[0.5mm]
IC~1613\dotfill		& 	7.78 $\pm$ 0.05			&		 \citet{Bresolin:2007a}				\\[0.5mm]
WLM\dotfill		& 	7.82 $\pm$ 0.08			&		 \citet{Lee:2005}					\\[0.5mm]
NGC~3109\dotfill	& 	7.79 $\pm$ 0.08			&		 \citet{Pena:2007}					\\[0.5mm]
NGC~6822\dotfill	& 	8.12 $\pm$ 0.07			&		 \citet{Lee:2006}					\\[0.5mm]
Sextans A\dotfill	& 	7.48 $\pm$ 0.02			&		 \citet{Kniazev:2005}					\\[0.5mm]
Sextans B\dotfill	& 	7.59 $\pm$ 0.20			&		 \citet{Kniazev:2005}					\\[0.5mm]
NGC~5253\dotfill	& 	8.20 $\pm$ 0.03			&		 \citet{Kobulnicky:1997, Lopez-Sanchez:2007}				
\enddata 
\end{deluxetable}


\subsection{Spiral galaxies with abundance gradients determined from \oiiiit\lin4363 detections.}
Four spiral galaxies with a well-determined oxygen abundance gradient based on the detection of \oiii\lin4363 are  presented in Table~\ref{table:spirals}.
 Cepheids in two fields at different galactocentric distances have been observed in M33 (\citealt{Scowcroft:2009}) and M101 \citet{Kennicutt:1998}. The abundances in the inner and outer fields of M33 and M101 were calculated from the following expressions for the abundance gradients:

\begin{trivlist}
\item {\em M33:} \oh\eq$8.42~ (\pm 0.03)~ -~ 0.26~ (\pm 0.07)~ R/R_{25}$, with $R_{25}$\eq35\farcm4 (\citealt{Bresolin:2010}). \\
{\phantom{\em M33:}} $(R_{inner} = 3\farcm6$, $R_{outer}=21\farcm1)$
\medskip

\item {\em M101:} \oh\eq$8.75~ (\pm 0.05)~ -~ 0.90~ (\pm 0.07)~ R/R_{25}$ , with $R_{25}$\eq14\farcm4 (\citealt{Bresolin:2007}).\\
{\phantom{\em M101:}} $(R_{inner} = 1\farcm7$, $R_{outer}=7\farcm9)$

\end{trivlist}
\medskip

The Cepheids studied in NGC~300 cover a wide range in radius, and one could crudely divide them into an inner and an outer sample,  using 
the following expression for the abundance gradient:

\begin{trivlist}

\item {\em NGC~300:} \oh\eq$8.57~ (\pm 0.02)~ -~ 0.41~ (\pm 0.03)~ R/R_{25}$ , with $R_{25}$\eq9\farcm75 (\citealt{Bresolin:2009a}).

\end{trivlist}

The oxygen abundance for NGC~300 in Table~\ref{table:spirals} is the average for \hii\ regions at  galactocentric distances larger than 0.3\,\rtf, where the bulk of the Cepheids is located.
In the case of NGC~2403 the abundances of the five \hii\ regions from \citet{Garnett:1997} at galactocentric distances larger than 2.7~kpc (0.25\,\rtf) were averaged. 
\bigskip

\begin{deluxetable}{lcl}
\tabletypesize{\scriptsize}
\tablecolumns{3}
\tablewidth{0pt}
\tablecaption{Cepheid host spiral galaxies with \oiii\lin4363 detections \label{table:spirals}}

\tablehead{
\colhead{\phantom{}Galaxy\phantom{aaaaaaaaaaaaaaa}}	     &
\colhead{\oh$_{Te}$}       &
\colhead{Sources of emission line data}  }
\startdata
\\[-2mm]
M33 (inner)\tablenotemark{1} \dotfill				& 	8.39 $\pm$ 0.04				&		 \citet{Bresolin:2010}								\\[0.5mm]	
M33 (outer)\tablenotemark{1}\dotfill				& 	8.26 $\pm$ 0.07				&		 \citet{Bresolin:2010}								\\[1mm]	
M101 (inner)\tablenotemark{2}\dotfill			& 	8.64 $\pm$ 0.06				&		 \citet{Kennicutt:2003, Bresolin:2007}				\\[0.5mm]
M101 (outer)\tablenotemark{2}\dotfill			& 	8.26 $\pm$ 0.08				&		 \citet{Kennicutt:2003, Bresolin:2007}				\\[1mm]
NGC~300\dotfill							& 	8.33$\pm$ 0.13				&		 \citet{Bresolin:2009a}							\\[1mm]	
NGC~2403\dotfill							& 	8.28$\pm$ 0.11				&		 \citet{Garnett:1997}								
\enddata 
\tablenotetext{1}{\citet{Scowcroft:2009}} \tablenotetext{2}{\citet{Kennicutt:1998}\\ \\ \\} 
\end{deluxetable}


\subsection{Oxygen abundances for Cepheid host galaxies using the transformation to the \te\ scale }
For galaxies in which no (or only a few) \oiii\lin4363 line detections are available we used Eq.~\ref{eq:zkh} to scale the oxygen abundances calculated in the \zkh\ scale to the \te\ scale. The results are presented in Table~\ref{table:all}.
The galaxy list was drawn from \citet{Sakai:2004}, \citet{Saha:2006} and \citet{Bono:2010}, with the addition of a few galaxies. Unless otherwise noted (see references at the end of the table), the abundances in the \zkh\ scale were taken from \citet{Ferrarese:2000}.
\bigskip

We comment below on a few galaxies from Table~\ref{table:all} in which auroral lines have been detected:

\begin{trivlist}

\item {\em NGC~224 (M31):} A value of \oh\eq$8.41\pm0.02$ has been reported by \citet{Esteban:2009} from a detection of \oiii\lin4363 in the outlying \hii\ region K932 ($R\simeq16$~kpc from the galaxy center). Given the rather shallow abundance gradient in this galaxy (still uncertain, but between $-0.006$ and $-0.027$ dex\,kpc\expone, \citealt{Trundle:2002}), 
this direct \te-based abundance is probably a good measurement of the overall metallicity. 

\item {\em NGC~1365:} \citet{Bresolin:2005} measured chemical abundances from the \nii\lin5755, \siii\lin6312 and \oii\lin7325 auroral lines for three \hii\ regions in the vicinities of the 
Cepheids studied by the {\em HST} Key Project. The weighted mean oxygen abundance  is \oh\eq$8.48\pm0.10$, in good agreement with the  value of 8.48 given in Table~\ref{table:all} for the \te\ scale.

\item {\em NGC~3031 (M81):} The detection of \oiii\lin4363 in \hii\ regions in the galactocentric radial range 5.8-9.8~kpc  has been reported by \citet{Stanghellini:2010}. The average abundance they derived is \oh\eq$8.54^{+0.18}_{-0.32}$. This metallicity should be representative of the inner field in Table~\ref{table:all}, which gives a lower value of 8.33.

\item {\em NGC~5236 (M83):} \citet{Bresolin:2005} measured chemical abundances from the \nii\lin5755, \siii\lin6312 and \oii\lin7325 auroral lines for two \hii\ regions in the vicinities of the Cepheids studied by \citet{Thim:2003}. The weighted mean oxygen abundance  is \oh\eq$8.55\pm0.10$, in good agreement with the value of 8.64 given in Table~\ref{table:all}.

\item {\em NGC~7793:} From the emission line data published by \citet{Edmunds:1984} for two \hii\ regions with good \oiii\lin4363 detection we derive \oh\eq$8.42\pm0.09$, in agreement with the 8.45 value in Table~\ref{table:all}.

\end{trivlist}

\begin{deluxetable}{lccc}
\tabletypesize{\scriptsize}
\tablecolumns{4}
\tablewidth{0pt}
\tablecaption{Metallicities of other Cepheid host  galaxies in the \te\ and \zkh\ scales \label{table:all}}

\tablehead{
\colhead{\phantom{}Galaxy\phantom{aaaaaaaaaaaaaaaaaaa}}	     &
\multicolumn{2}{c}{\oh} &
\colhead{error}  \\[0.5mm]
\colhead{}       &
\colhead{\phantom{aaaaa}\te\ scale \phantom{aaaaa}}   &
\colhead{\phantom{aaaaa}\zkh\ scale\phantom{aaaaa}}       &
\colhead{}   }
\startdata
\\[-2mm]
  NGC~55               					\dotfill &    8.06   &     8.35\tablenotemark{a}   &     0.15  \\[0.5mm]
  NGC~224              					\dotfill &    8.49   &     8.98   &     0.15  \\[0.5mm]
  NGC~925              					\dotfill &    8.19   &     8.55   &     0.15  \\[0.5mm]
  NGC~1309             					\dotfill &    8.44   &     8.90\tablenotemark{b}   &     0.19  \\[0.5mm]
  NGC~1326A         					\dotfill &    8.16   &     8.50   &     0.15  \\[0.5mm]
  NGC~1365             					\dotfill &    8.48   &     8.96   &     0.20  \\[0.5mm]
  NGC~1425             					\dotfill &    8.50   &     9.00   &     0.15  \\[0.5mm]
  NGC~1637             					\dotfill &    8.56   &     9.08\tablenotemark{c}   &     0.15  \\[0.5mm]
  NGC~2090             					\dotfill &    8.37   &     8.80   &     0.15  \\[0.5mm]
  NGC~2541             					\dotfill &    8.16   &     8.50   &     0.15  \\[0.5mm]
  NGC~2841             					\dotfill &    8.71   &     9.30\tablenotemark{d}   &     0.15  \\[0.5mm]
  NGC~3021             					\dotfill &    8.46   &     8.94\tablenotemark{b}   &     0.25  \\[0.5mm]
  NGC~3031 (inner)\tablenotemark{1}	\dotfill &    8.33   &     8.75   &     0.15  \\[0.5mm]
  NGC~3031 (outer)\tablenotemark{2}	\dotfill &    8.23   &     8.60\tablenotemark{e}   &     0.15  \\[0.5mm]
  NGC~3198             					\dotfill &    8.23   &     8.60   &     0.15  \\[0.5mm]
  NGC~3319             					\dotfill &    8.08   &     8.38   &     0.15  \\[0.5mm]
  NGC~3351             					\dotfill &    8.67   &     9.24   &     0.20  \\[0.5mm]
  NGC~3368             					\dotfill &    8.64   &     9.20   &     0.20  \\[0.5mm]
  NGC~3370             					\dotfill &    8.37   &     8.80\tablenotemark{f}    &     0.19  \\[0.5mm]
  NGC~3621             					\dotfill &    8.33   &     8.75   &     0.15  \\[0.5mm]
  NGC~3627             					\dotfill &    8.68   &     9.25   &     0.15  \\[0.5mm]
  NGC~3982            				 	\dotfill &    8.33   &     8.75\tablenotemark{f}   &     0.25  \\[0.5mm]
  NGC~4321             					\dotfill &    8.59   &     9.13   &     0.20  \\[0.5mm]
  NGC~4414             					\dotfill &    8.64   &     9.20   &     0.15  \\[0.5mm]
  NGC~4496A            					\dotfill &    8.35   &     8.77   &     0.15  \\[0.5mm]
  NGC~4527             					\dotfill &    8.33   &     8.75\tablenotemark{g}   &     0.15  \\[0.5mm]
  NGC~4535             					\dotfill &    8.64   &     9.20   &     0.15  \\[0.5mm]
  NGC~4536             					\dotfill &    8.40   &     8.85   &     0.15  \\[0.5mm]
  NGC~4548             					\dotfill &    8.74   &     9.34   &     0.15  \\[0.5mm]
  NGC~4639             					\dotfill &    8.50   &     9.00   &     0.15  \\[0.5mm]
  NGC~4725             					\dotfill &    8.45   &     8.92   &     0.15  \\[0.5mm]
  NGC~5236            	 				\dotfill &    8.64   &     9.20\tablenotemark{h}   &     0.15  \\[0.5mm]
  NGC~7331             					\dotfill &    8.28   &     8.67   &     0.15  \\[0.5mm]
  NGC~7793             					\dotfill &    8.45   &     8.92\tablenotemark{i}    &     0.12  \\[0.5mm]
  IC~4182              					\dotfill &    8.09   &     8.40   &     0.20  
\enddata 
\tablenotetext{1}{\citet{Freedman:1994}} \tablenotetext{2}{\citet{McCommas:2009}} 
\tablerefs{(a) \citet{Zaritsky:1994}; (b) \citet{Riess:2009a}; (c) \citet{Leonard:2003}; (d) using line fluxes from \citet{Bresolin:1999}; (e) \citet{McCommas:2009}
(f) \citet{Riess:2005}; (g) \citet{Saha:2006}; (h) using line fluxes from \citet{Bresolin:2005}; (i) using line fluxes from \citet{Webster:1983}    }
\end{deluxetable}

\end{document}